\documentclass[12pt]{article}

\usepackage{newtxtext,newtxmath}

\usepackage{graphicx}

\usepackage[letterpaper,margin=1in]{geometry}

\renewenvironment{abstract}
	{\quotation}
	{\endquotation}

\date{}

\makeatletter
\renewcommand{\fnum@figure}{\textbf{Figure \thefigure}}
\renewcommand{\fnum@table}{\textbf{Table \thetable}}
\makeatother

\usepackage{scicite}

\usepackage[hyphens]{url}

\def\scititle{
	Multivariate Joint Recurrence Quantification Analysis: detecting coupling between time series of different dimensionalities
}
\title{\bfseries \boldmath \scititle}

\author{
	Sebastian~Wallot$^{1\dagger}$,
	Dan~M{\o}nster$^{2\ast\dagger}$\and
	\small$^{1}$Institute for Sustainability  Psychology, Leuphana University of L{\"u}neburg, 21335 L{\"u}neburg, Germany.\and
	\small$^{2}$School of Business and Social Sciences, Aarhus University, 8000 Aarhus C, Denmark.\and
	\small$^\ast$Corresponding author. Email: danm@econ.au.dk\and
	\small$^\dagger$These authors contributed equally to this work.
}

\begin{document} 

\maketitle

\begin{abstract} \bfseries \boldmath
One challenge with the analysis of complex systems and the interaction between such systems is that they are composed of different numbers of components, or simply the fact that a different number of observables is available for each system. The challenge is how to analyze the interaction of two systems which are not described by the same number of variables. Here, we present multivariate joint recurrence quantification analysis (MvJRQA), a recurrence-based technique that allows to analyze coupling properties between multivariate datasets that differ in dimensionality (i.e., number of observables) and type of data (such as nominal or interval-scaled, for example). First, we introduce the methods, and test it on simulated data from linear and nonlinear systems. Then we apply it to an empirical dataset of EEG and eye tracking data. We introduce the joint recurrence coupling indicator (JRCI) as a measure to assess and compare coupling between systems. Finally, we discuss practical issues regarding the application of the method.
\end{abstract}

\noindent
The current paper presents an extension of extant recurrence analysis techniques, joint recurrence quantification analysis (JRQA \cite{marwan_recurrence_2007}) and multidimensional recurrence quantification analysis (MdRQA \cite{wallot_multidimensional_2016}). It is particularly aimed at quantifying the correlation---or coupling---between time series that differ in dimensionality, i.e., assessing coupling between two systems with a different number of observed variables. Examples include correlating a high-dimensional neurophysiological recording (e.g., electroencephalogram, functional magnetic resonance imaging, etc.) with a uni-dimensional behavior stream (e.g., gaze fixation times); the behavior of a group leader (e.g., uni-dimensional acceleration profile or transcript of speech)  with the collective behavior of the other group members (e.g., multidimensional acceleration profile or transcript of speech, where each group member adds one or more variables to the multivariate group dynamics); global climate models (temperature/pressure/humidity fields across grids, thousands of dimensions) vs. ice core isotopes (1D time series), i.e., correlating rich spatiotemporal data with a single low-dimensional paleoclimatic record.

The advantage of the method is that multidimensional signals do not need to be averaged or otherwise subjected to reduction procedures in order to reduce their dimensionality to one \cite{mayseless_real-life_2019}, or to the dimension of the lower-dimensional time series that they are to be correlated with. Applying such a reduction procedure is, in general, a requirement of correlational methods to have a matched set of paired data points in order to estimate correlation. In the approach presented here, the full dynamics of the multivariate time series are retained for analysis to produce their recurrence profile, which we describe in more detail below. This has particular advantages if the (multidimensional) time series in question exhibits complicated dynamics \cite{fuchs_coordination_2018,scholz_intentional_1990,ramenzoni_joint_2011,crone_synchronous_2021}, autocorrelation in terms of fractal fluctuations (a.k.a., long-memory, $1/f$ noise) \cite{stephen_dynamics_2009,kuznetsov_effects_2011,kello_pervasiveness_2008,he_scale-free_2011}, or inter-dependencies in terms of fractal correlations \cite{abney_complexity_2014,marmelat_strong_2012}. In particular, the occurrence of multifractal fluctuations in human behavioral and neurophysiological data suggests that different time series measured from a single organism show such interdependencies \cite{delignieres_multifractal_2016, ihlen_interaction-dominant_2010}. In all of these cases, fluctuations are informative about the behavior of the system and would get lost by averaging of the signals.

\begin{table}
	\centering
	\caption{\textbf{Examples of extant methods for assessing coupling between time series.}
		Many methods can be used to assess coupling between two uni-variate time series, whereas fewer methods are suited for multi-variate time series. Multivariate methods often rely on embedding time series in a phase space. MvJRQA is unique in allowing for time series of arbitrary dimensions.}
	\label{tab:methods}
    ~\\
    \includegraphics[width=0.9\textwidth]{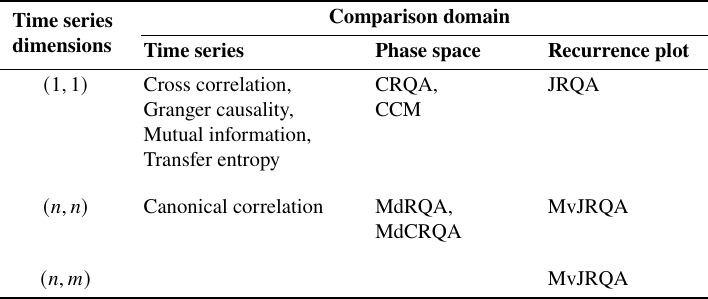}
\end{table}

Some of the current alternatives (see Table~\ref{tab:methods}) are various bi-variate analyses (cross-correlation, \cite{nelson-wong_application_2009}; Granger causality \cite{granger_investigating_1969}; transfer entropy \cite{schreiber_measuring_2000}; mutual information \cite{frenzel_partial_2007}; cross-recurrence \cite{shockley_cross_2002}; convergent cross mapping \cite{sugihara_detecting_2012}; and many others), which can also be arranged to provide a network-analytic portrait (e.g., \cite{paxton_network_2014}). However, such analyses only capture bi-variate relationships among the individual observables, but do not capture higher-level dynamics. Another approach is to reduce the dimensionality of the data, for example by simply averaging across all component time series. This, however, runs a risk of averaging out interesting dynamics from the data. The same goes for other techniques of dimension reduction, such as principal component analysis or factor analysis---which have the additional problem that the estimates of techniques are potentially vulnerable to auto-correlation properties that are usually present in time series data \cite{vanhatalo_impact_2016}. Finally, there are correlation methods that provide averages on the level of correlation parameters, such as canonical correlation \cite{thompson_canonical_1984} or multidimensional cross-recurrence quantification analysis \cite{wallot_multidimensional_2019}. While they are multivariate analyses that take inter-correlations among the different observables of multivariate time series into account, they always need a matching number of data points and variables. The method we propose here, Multivariate Joint Recurrence Quantification Analysis (MvJRQA), contains such analyses as a special case, but generalizes to situations where we seek to quantify correlation between two multivariate time series that do not have a matched number of observables. Moreover, since MvJRQA is a recurrence-based technique, it keeps more of the information inherent in variation in the data that are otherwise lost in the averaging process \cite{wallot_using_2013}. In this respect it is similar to the synchronization likelihood and time-dependent mutual information \cite{stam_synchronization_2002}, the joint probability of recurrence \cite{romano_detection_2005}, and the mean conditional probability of recurrence \cite{romano_estimation_2007}. Other recent work has introduced methods to infer causal network structure from multivariate time series \cite{runge_detecting_2019} and to infer unobserved causal drivers of multivariate time series \cite{gilpin_recurrences_2025}.

In the following we start by introducing MvJRQA and show the results of applying the method to four model systems that differ in dimensionality. Then, we show an exemplar empirical case, applying the method to simultaneous recordings of EEG and eye movement data. We finish by summarizing the results, the scope of the method in its current form, and provide some guidance for the application of the method.

\subsection*{Multivariate Joint Recurrence Quantification Analysis (MvJRQA)}
\begin{figure}
    \centering
    \includegraphics[width = 0.95\textwidth]{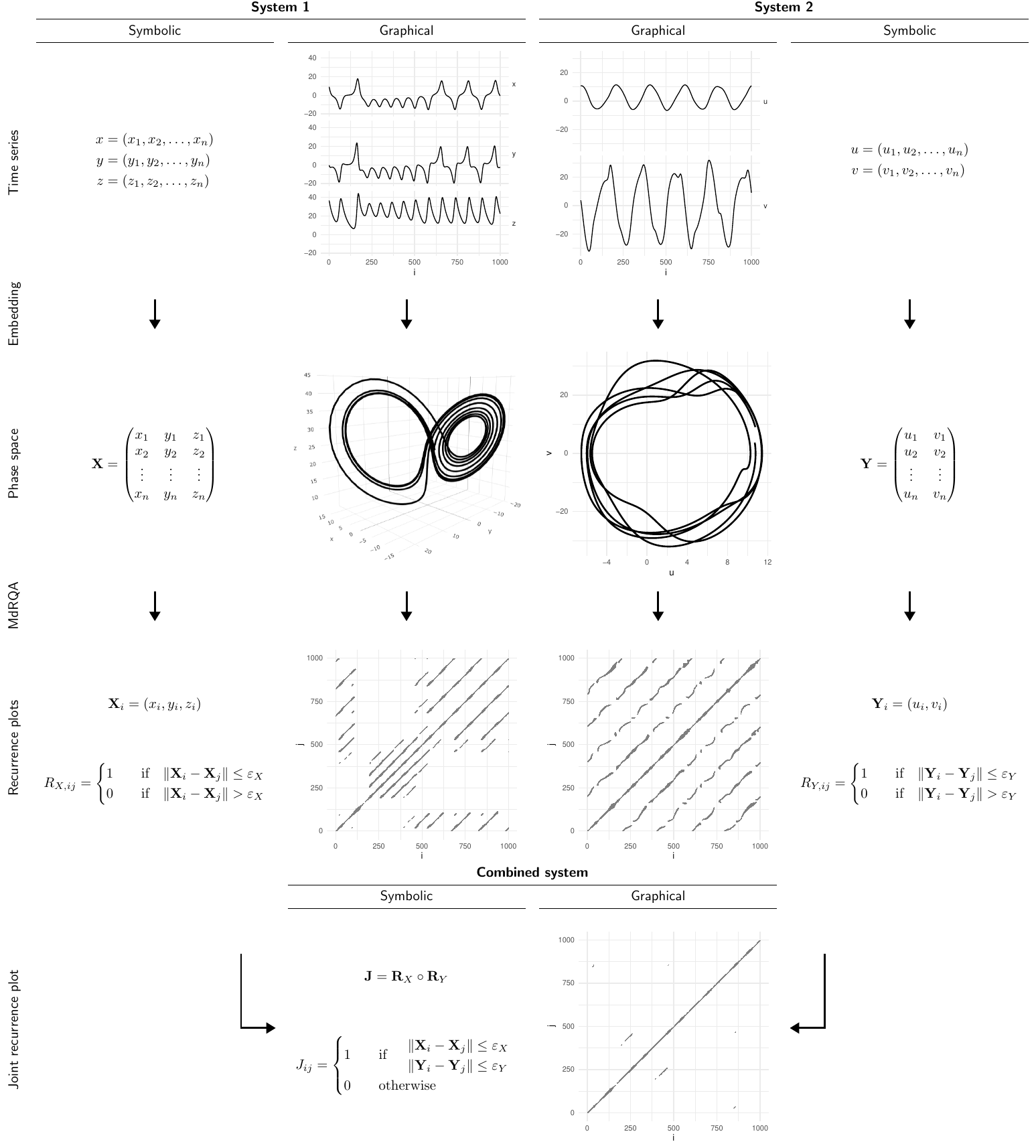}
    \caption{\textbf{Conceptual overview of constructing a multidimensional joint recurrence plot.} Two systems, shown in the columns `System 1' and `System 2,' have times series shown in the top row: \textit{Time series}. For each system a columns shows a symbolic representation, while the other column shows a graphical example. The first step, \textit{Embedding}, places the time series into a higher dimensional \textit{Phase space}, where the rows in $\mathbf{X}$ and $\mathbf{Y}$ represent points in phase space. Then \textit{MdRQA} is used to construct individual \textit{Recurrence plots} that are combined into a \textit{Joint recurrence plot}.}
    \label{fig:conceptual}
\end{figure}

MvJRQA extends Multidimensional Recurrence Quantification Analysis (MdRQA) \cite{wallot_multidimensional_2016} by combining it with Joint Recurrence Plots (JRPs) \cite{romano_multivariate_2004}, a method to extract measures of similarity between two time series. The idea behind MvJRQA is that if two systems---or two subsystems of a larger system---are coupled to each other, then this coupling will affect the dynamics of the systems in ways that can be quantified by looking at simultaneous recurrences (\emph{i.e.,} joint recurrences) of the two systems. In this section, we will briefly explain MdRQA and JRP as well as how we combine these methods to obtain MvJRQA. A conceptual overview of the method is provided in Figure~\ref{fig:conceptual} where a 3-dimensional system (the Lorenz system) is coupled to a 2-dimensional system (a harmonic oscillator).
First, we start with datasets from two systems (columns `System 1'  and `System 2' in Figure~\ref{fig:conceptual}) we know or hypothesize to be coupled through some interaction. One system is three-dimensional (system 1), while the other system is two-dimensional (system 2). In other words, we have three observables from system 1 and two from system 2 (row `Time series' in Figure~\ref{fig:conceptual}). Next, the phase-space trajectories of the two systems are obtained by embedding the components in a phase space, or visually: plotting the observables against each other in a single phase space (row `Phase space' in Figure~\ref{fig:conceptual}). In our example this is a three-dimensional space for the three time series form system 1, and a two-dimensional space for the two time series from system 2. In the more general case the time series can be embedded into a higher dimensional space via the method of time-delayed embedding, introduced in the next section. Then, we compute individual recurrence plots (explained below) for each of the two system's sets of time series using MdRQA (row `Recurrence plots' in Figure~\ref{fig:conceptual}) and finally join these two plots (JRP) to get a recurrence matrix of their shared dynamics, which is obtained by the element-wise multiplication of the two recurrence matrices (row `Joint recurrence plot' in Figure~\ref{fig:conceptual}). From this joint recurrence plot, measures can be extracted that capture the properties of their shared dynamics, as well as their coupling strength \cite{marwan_recurrence_2007}. We will go through the details of analyzing these coupled systems in the results, as well as materials and methods sections.

\subsubsection*{Recurrence plot and joint recurrence plot}
For the purpose of illustration, we assume here that we are dealing with a time series, i.e., a set of measurements at regular intervals of a single variable $x$ to give a set of values $(x_1, x_2, \ldots, x_n)$. We further assume that $x$ is one of several variables needed to fully describe some system, and that an approximate description of the full system can be obtained by the method of time-delayed embedding \cite{takens_detecting_1981}, resulting in the construction of a multidimensional phase space embedding of the uni-dimensional variable $x$. If $x$ is embedded into an $m$-dimensional space with time delay $\tau$ the points will be vectors of the form
\begin{equation}
    \label{eq:embedding}
    \mathbf{X_i} = (x_i, x_{i - \tau}, x_{i - 2\tau}, \ldots, x_{i - (m-1)\tau}).
\end{equation}
The embedding parameters $m$ and $\tau$ are not given a priori, but must be estimated from the data \cite{wallot_calculation_2018}.

The fundamental concept of all recurrence-based methods is the recurrence plot \cite{eckmann_recurrence_1987} which is a graphical visualization of when a system's state recurs, i.e., if the system is in a state $\mathbf{X}_1 = (X_1, X_2, \ldots, X_m)$ at time $t_1$ and again at time $t_2$, the points $(t_1, t_2)$ and $(t_2, t_1)$ will be included in the recurrence plot of $x$. We can formulate this mathematically in terms of the recurrence matrix $\mathbf{R}$ with elements
\begin{equation}\label{eq:RR_def}
    R_{ij} = 
    \begin{cases} 
      1 & \quad \text{if}\quad \|\mathbf{X}_i - \mathbf{X}_j\| \leq \varepsilon \\
      0 & \quad \text{if}\quad \| \mathbf{X}_i - \mathbf{X}_j \| > \varepsilon.
   \end{cases}
\end{equation}
Here, $\|\cdot\|$ is a norm in the $m$-dimensional phase space (usually the Euclidean distance) and $\varepsilon$ is a small radius within which two points will be considered equal and therefore recurrent (see \cite{marwan_recurrence_2007} for a comprehensive introduction to recurrence plots).

A joint recurrence plot (JRP) is an element-wise product of two separate recurrence plots (recurrence matrices) with the same dimensions. This means that a JRP has elements with value 1 when \emph{both} of the individual RP's have the value 1 for a particular pair of times $(t_1, t_2)$. If we have the time series $x$ and another time series $y$ with phase space coordinates $\mathbf{X}_i$ and $\mathbf{Y}_i$ (constructed according to Equation~\ref{eq:embedding}) then we can define the joint recurrence matrix $\mathbf{J}$ by its elements
\begin{equation}\label{eq:J}
    J_{ij} = 
    \begin{cases} 
      1 & \quad \text{if}\quad \|\mathbf{X}_i - \mathbf{X}_j\| \leq \varepsilon_x \quad\text{and}\quad \|\mathbf{Y}_i - \mathbf{Y}_j\| \leq \varepsilon_y \\
      0 & \quad \text{otherwise}.
   \end{cases}
\end{equation}
Here $\varepsilon_x$ and $\varepsilon_y$ are the radius parameters defining recurrences of $x$ and $y$, respectively. If $x$ has the recurrence matrix $\mathbf{R}_x$ and $y$ has the recurrence matrix $\mathbf{R}_y$ then the above can also be written $\mathbf{J} = \mathbf{R}_x \circ \mathbf{R}_y$, where ``$\circ$'' denotes the element-wise product of two matrices.

\subsubsection*{Multidimensional extension to recurrence analysis}
Multidimensional Recurrence Quantification Analysis  \cite{wallot_multidimensional_2016} is an extension of Recurrence Quantification Analysis \cite{webber_dynamical_1994} where an inherently multidimensional time series can be analyzed. It also allows multiple uni-dimensional time series---such as data from a group of interacting individuals---to be aggregated into a multidimensional time series, thus facilitating the analysis of data from groups larger than dyads.\footnote{Dyadic data can be analyzed using the bivariate method Cross Recurrence Quantification Analysis \cite{zbilut_detecting_1998}.} So, while RQA quantifies the dynamics of a univariate time series, MdRQA quantifies the dynamics of a multivariate time series which may be constructed from several univariate time series considered to be parts of a bigger dynamical system. It can be viewed as a generalized time-dependent multivariate correlation measure, which along with the relevant multidimensional parameter estimation methods \cite{wallot_calculation_2018} are readily available through the R package \textsc{crqa} \cite{coco_r_2021}. All analyses in the current manuscript are done using this package together with a wrapper function to conduct MvJRQA which is available online \cite{monster_mvjrqa_2025}.

\subsubsection*{Multivariate joint recurrence analysis}
We now combine JRP and MdRQA to construct the method of Multivariate\footnote{For MdRQA we used the term ``multidimensional'' to highlight the multidimensional nature of the trajectory in phase space. For MvJRQA we have chosen ``multivariate'' to highlight that the method takes  multiple time series whose dynamics are combined to yield statistical outcome variables. However, the joint recurrence plot from which these outcome variables are computed, remains two-dimensional.} Joint Recurrence Quantification Analysis (MvJRQA). The method is applicable to situations where there are two interacting systems or subsystems, where one or both are best described using a multivariate time series. In essence, the method is to first use the techniques of MdRQA to construct a recurrence plot (i.e., a recurrence matrix) for each of the two systems based on the multivariate time series. The joint recurrence plot is then constructed as the element-wise product of the recurrence matrices for the two systems (cf.\ Figure~\ref{fig:conceptual}). Because MdRQA takes any number of time series to construct a recurrence plot of their dynamics, and because these recurrence plots can be joined, no matter how many component time series go into the composition of each individual recurrence plot using the MdRQA routine, this allows to quantify the joined dynamics of systems that differ in their number of observables.
There are multiple outcome variables that can be computed from recurrence plots to quantify the (shared) dynamics of time series \cite{marwan_recurrence_2007}. Here, we focus on the quantification of correlation (or coupling) between two systems and use the most basic property of a recurrence plot, which is percent recurrence ($RR$). Percent recurrence is computed as the sum of all recurring points (i.e., black dots) in a recurrence plot divided by the size of the recurrence plot:

\begin{equation}
JRR = 100\% \cdot \frac{1}{N^2} \sum_{i=1}^{N} \sum_{j=1}^{N} J_{ij}
\end{equation}

With $J$ being the joint recurrence plot, $N^2$ the size of the plot (i.e., the number of elements in the $N \times N$ recurrence-matrix). Everything else equal, two (sets of) time series that exhibit similar dynamics will also result in similar recurrence profiles (i.e., distributions of recurrence points on their respective recurrence plots). If their plots are joined as described above, this will result in higher rates of joint recurrences (i.e.,  $JRR$ of the joint recurrence plot). Hence, from $JRR$ of the joint recurrence plot, we can derive measures of coupling strength to quantify coupling between different systems (see Results section).

Finally, the values of the radius parameters, $\varepsilon_x$ and $\varepsilon_y$, in Equation~\ref{eq:J} will be of great importance for the method: First, the radius parameters can be adjusted to increase or decrease the recurrence rate $RR$ of the individual RPs to be joined. The bigger the radius chosen, the more data points will count as recurrent. Hence, we can use the radius parameters to adjust the resulting recurrence rates $RR$ for the individual plots to be joined.
Second, the joint recurrence rate, $JRR$, depends on the recurrence rates $RR$ of the individual RPs to be joined, with the maximum $JRR$ limited by the minimum $RR$ of the two individual plots. Hence, if one of the individual plots has low $RR$, this implies that the $JRR$ will also be low, no matter how similar the two multidimensional time series are.
Accordingly, it is important to adjust the radius parameter for the individual RPs to fix the recurrence rates $RR$ of the individual plots to the same level, so that potential differences in $RR$ between the two individual RPs will not impact the results of the analysis.

\subsection*{Results}
\begin{figure}
    \centering
    \includegraphics[width=0.99\linewidth]{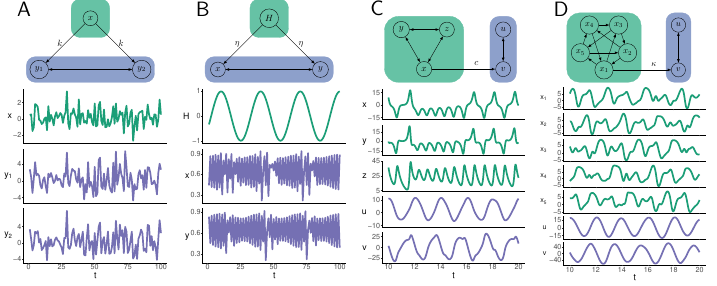}
    \caption{\textbf{Model systems and example time series.} A linear system with one stochastic process coupled to two correlated stochastic processes via weight $k$ (\textbf{A}). A periodic signal driving a set of two coupled logistic maps via coupling parameter $\eta$ (\textbf{B}). The canonical Lorenz system driving a harmonic oscillator via a coupling parameter $c$ (\textbf{C}). The Lorenz-96 system with 5 dimensions driving a harmonic oscillator via a coupling parameter $\kappa$ (\textbf{C}). Colors indicate the driving (green) and driven (purple) systems.}
    \label{fig:model-timeseries}
\end{figure}

Here, we first present results from the application of MvJRQA to three different model systems. They represent different possible applications of MvJRQA to systems exhibiting nonlinear dynamics or simple stochastic system. Figure~\ref{fig:model-timeseries} provides an overview over the model systems. The details of the model systems are described in the Methods section. Moreover, we apply MvJRQA to empirical data comprised of co-registered eye movements and EEG during surgery simulations, that differ in terms of their dimensionality, but are expected to be coupled in terms of their dynamics. 

\subsection*{Model systems}
To determine how well MvJRQA is able to detect coupling for linear stochastic systems, as well as systems where nonlinear, time-dependent dynamics are involved, we investigate four sets of model systems where all interactions are known and can be varied. We have included both continuous time systems (flows) and discrete time systems (maps), with both linear and nonlinear components, stochastic and deterministic systems, as well as systems with different number of dimensions. This provides a diverse set of model systems to validate the ability of MvJRQA to detect coupling. The first set of systems is a linear coupling of simple stochastic systems. Of course, random variables that are simply linearly combined and do not possess complex dynamics, and other modeling alternative exists for such cases, such as latent variable modeling (e.g., \cite{mcardle_latent_2009}). The point here is to show that the MvJRQA procedure can also be used to recover such effects for simple linear systems with stochastic data.

The second set of systems is two coupled logistic maps \cite{may_simple_1976} in the chaotic regime driven by a periodic external signal. We use a simple cosine as the external signal, which has been used to represent cyclical variation in environmental conditions in ecosystems models \cite{summers_chaos_2000, monster_causal_2017}. A single, isolated, logistic map is either stationary or displays periodic or chaotic dynamics, depending on the growth rate. We include the coupled logistic maps with external driver as an example of a discrete-time system with very complex dynamics.

The last two sets of systems are the the canonical Lorenz system and the Lorenz-96 system (both of which are nonlinear systems), each driving a harmonic oscillator.
The canonical Lorenz system is characterized by chaotic behavior, a strange attractor and sensitive dependence on initial conditions. It was introduced by Lorenz as a simplified model for convection in the atmosphere \cite{lorenz_deterministic_1963}. The Lorenz-96 system was introduced in 1996 and can vary in its dimensionality \cite{lorenzPredictability2006}. The harmonic oscillator describes the dynamics of a system under a force that is proportional to the displacement from the static equilibrium state. This can describe, \textit{e.g.,} a mass on a spring, a pendulum with small amplitude motion, certain electrical circuits as well as neurological and physiological oscillations. The harmonic oscillator displays predictable periodic dynamics, and the attractor is a circle or ellipse.

\begin{figure}
    \centering
    \includegraphics[width=0.99\linewidth]{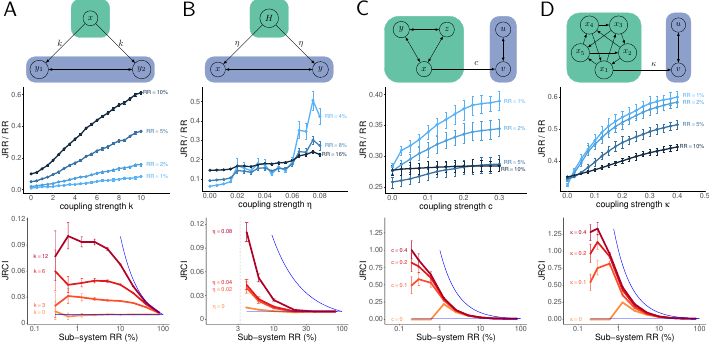}
    \caption{\textbf{Overview of results for the model systems.} \textit{Top row:} model systems as in Figure~\ref{fig:model-timeseries}. For each value of coupling strength, 100 independent models were run with different initial conditions. \textit{Middle row:} $JRR/RR$ is seen to increase monotonically with coupling between systems for a range of fixed subsystem $RR$. \textit{Bottom row:} Plots of JRCI for different values of the coupling show ability to correctly order systems by coupling strength with weakly interacting systems closer to the random null model and non-interacting systems consistent with the null model for low $RR$. Error bars indicate bootstrap estimates of 95\% confidence intervals. }
    \label{fig:results-overview}
\end{figure}

To manipulate coupling strength, we varied a coupling parameter linking the two systems of each set from zero (i.e., no coupling) to some number specific to each system that constituted extreme coupling (see Materials and Methods section for details). As described above, higher coupling will lead to higher $JRR$ for the joint recurrence plot of the two (sets of) time series, indicative of coupling.
However, in order to meaningfully compare $JRR$ across coupling conditions, it is important to keep the recurrence rate of the individual subsystems fixed. Otherwise a change in subsystem recurrence rate could be driving the change in joint recurrences (that is, if the individual RPs to be joined have high recurrence rate, they will produce a lot of joint recurrence simply as a function of their high base-rate of $RR$). Keeping recurrence rate fixed can be achieved by tuning the radius parameter $\varepsilon$ so that each set of time series from the two systems yield the same overall percent recurrence ($RR$). Moreover, the joint percentage recurrence can be normalized by the average percentage recurrence of the recurrence plots of the individual subsystems ($JRR/RR$)---also referred to as the synchronization index, $S$ [see \cite{marwan_recurrence_2007}, p.~292]. Note that here we define $RR$ to be the average individual subsystem recurrence rate, i.e., $RR = (RR_1 + RR_2)/2$, and we strive for almost equal recurrence rate of the two systems: $RR_1 \approx RR_2$.
This ratio, $JRR/RR$, will also improve interpretation of the results when percent recurrence of the individual recurrence plots cannot be fixed to exactly the same percentage.
The results are summarized in Figure~\ref{fig:results-overview}.
For ease-of-comprehension, the upper row presents the three model systems again.
The middle row of panels displays how $JRR/RR$ changes in the four model systems as a function of coupling strength with different fixed percentages of recurrence ($RR$) for individual subsystems. First of all, it is evident that $JRR/RR$ increases for the linear model (A) and the Lorenz systems and harmonic oscillator models (C and D) with increasing coupling strength, showing that $JRR/RR$ reliably distinguishes between coupling levels.
Here, we also see that it plays a role whether subsystem $RR$ is fixed at a relatively high vs. relatively low value: In general, the best monotonically increasing functions of $JRR/RR$ are obtained by individual subsystem $RR$ in the range 1\% to 5\% for the linear stochastic systems (A) and the two nonlinear systems driving a harmonic oscillator model (C and D).
For the coupled logistic maps driven by a harmonic signal (B) $JRR/RR$ also increases with coupling strength, but the curves appear less regular and are not always monotonically increasing. This system has notoriously complex dynamics which has also challenged other methods to detect coupling such as convergent cross mapping \cite{monster_causal_2017}. Nevertheless, the general trend is for synchrony to increase with subsystem recurrence rate. We note that for this system the lowest possible value of $RR$ is determined by the period of the harmonic signal $H(t)$ and with the chosen period, this is $RR \approx 3\%$, as indicated by the vertical dashed line in the bottom plot in panel B. 

In the limit $RR = 100\%$, $JRR/RR$ is trivially equal to one. This is also true for two systems with identical dynamics, where $JRR = RR$ and hence $JRR/RR$ also equal one. So, this ratio is generally uninformative with high $RR$ close to $100\%$, while it suggests strong (or perfect) coupling when subsystem $RR$ is low. Hence, the ability of one subsystem to perturb and change the trajectory of another subsystem can therefore be detected in the recurrences at low recurrence rate. If the systems are coupled, their interaction is likely to result in an increase in joint recurrences. Accordingly, we are primarily interested in joint recurrences at low subsystem $RR$, since the higher the $RR$ the higher the probability of chance (or spurious) joint recurrences is.

In the case of two independent stochastic systems joint recurrences occur only by chance, i.e., with a probability $P_r = RR^2$, where the subscript $r$ denotes a random joint recurrence. The joint recurrence rate is therefore $JRR_r = RR^2$. We refer to this as the \emph{random null model}. In the case of two identical systems every recurrence is also a joint recurrence, so the joint recurrence rate is equal to the subsystem recurrence rate: $RR_i = RR$, where the subscript $i$ denotes identical systems, and we refer to this as the \emph{identical systems model}. This is the theoretical maximum of $JRR$ and represents the limiting case approached by two nearly identical systems or, in the case of very strong coupling, two nearly synchronous systems [see \cite{sm_methods} for details]. A plot of $JRR/RR$ vs.\ $RR$ will therefore be delimited by these two cases where $JRR_r$ is a straight line from 0 at $RR = 0$ to 1 at $RR = 100\%$ and $JRR_i$ is a horizontal line at $JRR_i/RR = 1$ for all values of $RR$. However, since recurrence plots are usually constructed to be rather sparse, e.g. $RR < 10\%$, we are mostly interested in the behavior at low values of $RR$, so it is more convenient to plot $JRR/RR^2$, i.e., we divide by $RR^2$, rather than $RR$. In such plots $JRR_r/RR^2$ will now be a horizontal line at $JRR/RR^2 = 1$, whereas $JRR_i/RR^2 = 1/RR$.

Thus, in order to be able to discern the differences at low values of subsystem RR, we divide the relative joint recurrence rate $JRR/RR$ by another factor RR to obtain $JRR/RR^2$ which we refer to as the Joint Recurrence Coupling Indicator (JRCI). This construction of JRCI increases the sensitivity to detect close recurrences over more distant or less exact recurrences as $\varepsilon$ is increased in order to increase $RR$. This is not just a pragmatic choice---we also expect, theoretically, that the joint recurrences in the low $RR$ limit are more informative about coupling. To obtain a low, fixed, $RR$ for both systems a sufficiently small phase space radius, $\varepsilon$ is required (cf.\ Equation~\ref{eq:RR_def} and \ref{eq:J}). If coupling is not so strong that the two systems are completely synchronized the recurrences will mainly be determined by each systems intrinsic dynamics. As $RR$ is increased, by increasing $\varepsilon_x$ and $\varepsilon_y$, additional points in phase space will be recurrent for each system, and if the two systems are coupled the shared dynamics will make it more likely that some of these recurrences are joint recurrences. Therefore the fraction of joint recurrences relative to the fixed recurrence rate of the two systems is indicative of the coupling between the two systems, and in order to compare this fraction for different values of $RR$, we normalize by dividing with $RR$, thus obtaining JRCI. As $RR$ becomes large many recurrences will be spurious, i.e., they will only be counted as recurrences because $\varepsilon$ is very large. It is not possible to determine a priori which $RR$ values are relevant, unless we already know the details we are trying to infer, so we plot JRCI for a range of $RR$ values. 

The bottom row in in Figure~\ref{fig:results-overview} displays JRCI against subsystem RR for different coupling strengths. Also shown is JRCI for the random null model (lower horizontal blue line) and the identical system model (upper blue curve). 
The lowest JRCI-curve, corresponding to no coupling $(c = k = \eta = 0)$ could naively be expected to coincide with the blue line for the random null model corresponding to two independent stochastic systems. However, in the cases (C and D) of the coupled Lorenz systems and harmonic oscillator, the data are \emph{not} stochastic and therefore do not produce recurrence plots with points that are distributed according to a uniformly random distribution. Instead both subsystems display cyclic behavior and such commonalities in the dynamics will also lead to joint recurrences even in the absence of any coupling, as evident by the `bump' for $c = 0$ when subsystem $RR$ is in the range 1\% to 9\%. However, even in this case JRCI for zero coupling is below the values observed for even weak coupling $c = 0.1$ and $\kappa = 0.1$.

From the four model systems, the following conclusions can be drawn regarding the sensitivity of MvJRQA to detect coupling and the choice of fixing subsystem $RR$ at certain levels:  For the linear stochastic system (A), the ability of MvJRQA to distinguish the different coupling conditions is not very sensitive to  the choice of subsystem $RR$ (except for $RR \to 100\%$, of course). Even for subsystem $RR$ of 80\% (not shown on the graph), the analysis remains sensitive, even though it is most sensitive for subsystem $RR$ of 1\% to 20\%, when comparing the difference in average JRCI between the coupling conditions to the size of the 95\% CIs.

For the externally driven logistic maps (B), MvJRQA remains sensitive across the range of subsystem $RR$ from 4\% to 20\%, almost equally so across the whole range of subsystem $RR$ values. However, at subsystem $RR$ of 40\% and higher (not shown on the graph), the analysis looses its capacity to detect coupling, because average JRCI for the uncoupled condition now starts to increase above JRCI of the coupled conditions.

For the canonical Lorenz and harmonic oscillator system (C), the average JRCI of the uncoupled systems becomes indistinguishable from the average JRCI of the weak coupling condition ($c = 0.1$) when fixing $RR$ of the subsystems to 10\%. When fixing subsystem $RR$ to 20\%, the average JRCI of all three coupling conditions ($c = 0.1$ to $0.4$) lie within the 95\% CI of the JRCI of the uncoupled condition ($c = 0.0$). At this point, MvJRQA does not distinguish the different coupling conditions anymore. When comparing the difference in average JRCI between the coupling conditions to the size of the 95\% CIs, the analysis is most sensitive when fixing subsystem $RR$ to levels between 1 and 5 percent.

For the Lorenz~96 and harmonic oscillator system (D), adjacent coupling levels (i.e., 0 vs. 0.1, 0.1 vs 0.2 etc) become indistinguishable, when subsystem $RR$ is fixed to at about 20\%. When subsystem $RR$ approaches 40\%, none of the coupling levels are distinguishable from each other anymore. Again, the analysis is most sensitive for subsystem $RR$ between 1 and 5\%. Note that these results are stable even if the number of dimensions of the Lorenz~96 system is substantially increase (i.e., to 16---see figure~\ref{fig:lor96hidim}). Accordingly, MvJRQA does not seem to be affected by mere changes in the dimensionality of to-be-compared systems.

In general, we can say that MvJRQA performs consistently well across the model systems when subsystem $RR$ if fixed to values between 1 and 5 percent as a general recommendation, even though there are individual deviations from this rule, where the analysis also works well for higher subsystem $RR$ (as in the case of the linear stochastic systems, for example).

\subsubsection*{Comparison to MdRQA}

An analysis that captures coupling between systems of different dimensionality can, in principle, also be done using Multidimensional Recurrence Quantification Analysis (MdRQA; \cite{wallot_multidimensional_2016}). To do such an analysis, we simply take the observable from one system and the observable from the other system, and embed them into the same phase space (effectively treating the data as being from a single system).

Let us examine how MdRQA compares to MvJRQA for the three model systems. For Multivariate Joint Recurrence Quantification Analysis, we perform the same analysis as described above. For Multidimensional Recurrence Quantification Analysis, the procedure is a little different. First, we embed all time series into a single phase space for each of the model pairs.  Then, we set a single value for the threshold parameter $\varepsilon$, which we keep constant across all iterations over the range of the coupling parameters. We need to keep $\varepsilon$ constant, so that time series with stronger coupling can yield higher recurrence rates compared to time series with weaker coupling, given that we do not change the threshold of which values we count as recurrent. Furthermore, we normalized all data by z-scoring them, so that that changes in recurrence rates across iterations are due to differences in the sequential properties of the data, and not due to differences in variance of individual time series. 

\begin{figure}
    \centering
    \includegraphics[width=0.9\linewidth]{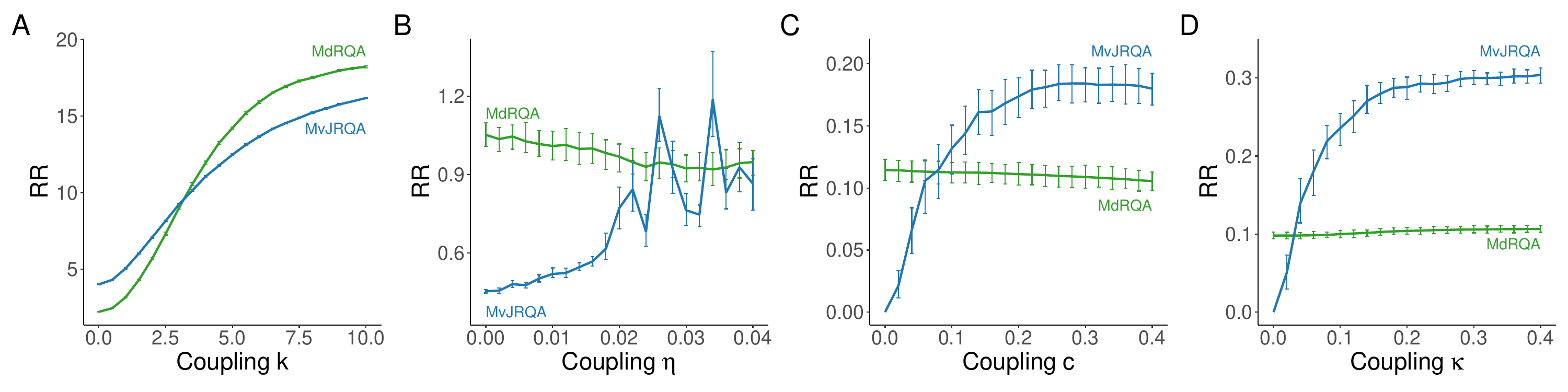}
    \caption{\textbf{MvJRQA compared to MdRQA for the  model systems.} For each value of coupling strength, 100 independent models were run with different initial conditions to generate a time series of length 500. The error bars indicate bootstrap estimates of 95\% confidence intervals. For the linear system (A), both, MvJRQA and MdRQA show increased $JRR$ or $RR$ with increased coupling. However, MdRQA does not pick up coupling in $RR$ for the three nonlinear systems (B--D).}
    \label{fig:compare-mdrqa-coupling}
\end{figure}

As can be seen in Figure~\ref{fig:compare-mdrqa-coupling}, recurrence rates for MdRQA increase with coupling for the linear system (A), but is not sensitive to increases in coupling for the other three systems (B-D), which involve nonlinearities. If anything, there is a slight tendency for $RR$ to decrease. Hence, this suggests that MvJRQA is more sensitive to detect differences in coupling across a wider range of systems with different properties.

\subsubsection*{Extreme coupling}

\begin{figure}
    \centering
    \includegraphics[width=0.99\linewidth]{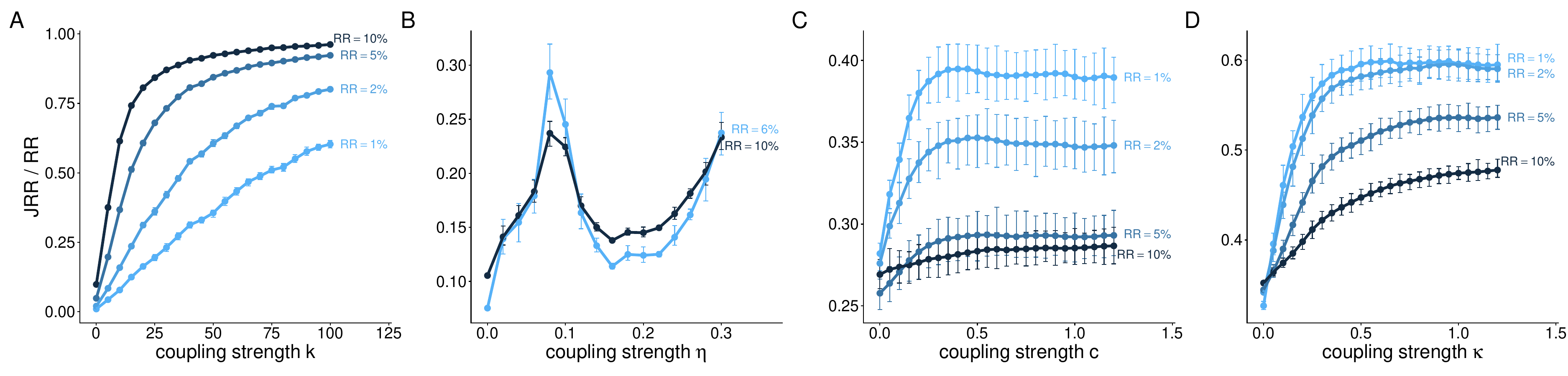}
    \caption{\textbf{Extreme coupling.} As a method to detect coupling in sets of time series, MvJRQA is compromised in situations of extreme coupling. For models A, C, and D we see a leveling-off of $JRR/RR$ with extreme coupling values, which is somewhat less pronounced for the linear systems (\textbf{A}), particularly when fixed $RR$ is small. For the coupled logistic maps system (B), extreme coupling leads to a significant drop and renewed increase in $JRR/RR$ with increase of coupling strength to extreme values.}
    \label{fig:extreme-coupling}
\end{figure}

It is a common challenge for analysis methods that are used to detect coupling to get correct results in cases where coupling becomes extreme \cite{monster_causal_2017, bresar_directional_2023}. The same is true for MvJRQA. One reason is that, theoretically, relative recurrence loses the ability to detect changes in coupling strengths in the case of extreme coupling. When coupling becomes so extreme that one time series comes to dominate the other time series so much that the driven time series starts to be almost identical to the driving time series, then this implies that the two recurrence plots of the individual time series are almost identical as well.
This in turn leads to a joint recurrence plot of the two individual time series' RPs, which is again very similar to the two individual RPs. If this is the case, but coupling is increased even further, it cannot, however, lead to substantive increases in joint recurrence anymore, because the JRP is almost identical to the two individual RPs already.
We performed further simulations to investigate the behavior of MvJRQA for cases of extreme coupling for our three model systems. As can be seen in Figure~\ref{fig:extreme-coupling}, this expectation is borne out for the linear system (A) and the Lorenz systems (C and D) driving the harmonic oscillator. However, a different pattern emerges for the logistic map (B). Here, MvJRQA fails in a different way to handle extreme coupling. Instead of reaching a plateau, the $JRR/RR$ ratio drops again in the regime of extreme coupling. A similar effect has been observed for the coupled logistic map system (without the external driver) in an application of convergent cross mapping, where the inferred coupling strength changes non-monotonically as a function of the actual coupling parameter \cite{monster_causal_2017}. 

\subsection*{Eye movement and EEG data}

Finally, we apply MvJRQA to an empirical dataset of co-registered eye movements and EEG. The dataset \cite{shafiei_integration_2023} was obtained from Physionet \cite{goldberger_physiobank_2000} and the data were recorded during practice tasks for robotic surgery from participants with different experience levels (from pre-medical students to faculty). Details of the study can be found in the original publication \cite{shafiei_performance_2023}. Here, we are primarily interested in using these data as a proof-of-concept for MvJRQA as a method for investigating coupling between dynamic processes of different dimensionality---here, EEG and eye movements.

Specifically, the dataset contains recordings of 2D eye movements and projections of eye movements into 3D. For at least some of the robotic surgery training tasks, gaze in three dimensions is necessary in order to perform the task well. Accordingly, it seems plausible that 3D eye movements will yield a stronger coupling relationship to EEG records compared to 2D eye movement data.

The dataset consists of co-registered 128-channel EEG, as well as 2D and 3D eye movements for the 25 participants of the original study \cite{shafiei_performance_2023}. Prior to computing MvJRQA, we removed four channels consisting of EOG data from the EEG data, since these channels measure the orientation of each eye's electrical dipole. Hence, including these would introduce the eye gaze into the EEG signal and would include the eye gaze in both subsystems. An example of the remaining 124 EEG channels and 2D eye tracking time series for a single trial from the dataset is shown in figure~\ref{fig:empirical-timeseries}. The original dataset contains 315 experimental trials each consisting of EEG data, eye tracking data and performance data. We removed two trials where the eye tracking data was missing and an additional seven trials because of data processing issues (see Materials and Methods section). We therefore ended up with 306 trials, giving a total of 612 observations, since each trial contains both 2D and 3D eye tracking data.

To assess our proposal that 3D eye movements show stronger coupling to EEG compared to 2D eye movements, because the specific nature of some of the surgery tasks requires visual attention in three dimensions, we estimated a mixed linear model with JRCI as the dependent variables ($y$), and eye movement signal dimensionality (EyeType: 2D or 3D) as fixed predictor variable and random intercepts $u_{0j}$ for participants:

\begin{equation}
\mathbf{y}_{ij} = \gamma_{00} + \gamma_{11}\text{EyeType}_{ij} + u_{0j} + r_{ij},
\label{eq:mlm}
\end{equation}
where $i$ indicates the repeated measurement within participants, and $j$ indicates the participant.

\begin{table}
\caption{\textbf{Regression results for MvJRQA.} The table shows fixed effects as standardized regression coefficients for the regression with JRCI as the dependent variable computed for EEG coupled with either 2D or 3D eye movement dynamics.  Subsystem recurrence was fixed at $RR = 2\%$ and 169 observations were removed because the recurrence rate differed from the target rate by more than 1.5 percentage points.}
\begin{center}
\begin{tabular}{l c c c l}
\hline
 & Estimate & SE & $t$ & $p$ \\
\hline
(Intercept)    & $-0.27$           & $0.10$ & $-2.61$ & $0.009$ \\
               & $ [-0.47; -0.07]$ & $$     & $$      & $$     \\
EyeType3D      & $0.34$            & $0.10$ & $3.33$  & $0.001$ \\
               & $ [ 0.14;  0.54]$ & $$     & $$      & $$     \\
\hline
Num. obs.      & $443$             & $$     & $$      & $$     \\
Conditional $R^2$ & $0.10$            & $$     & $$      & $$     \\
\hline
\multicolumn{5}{p{0.6\linewidth}}{\small{Note: EyeType is a dummy variable that codes for whether 2D or 3D eye movements were used together with the EEG data. Hence, the predictor EyeType3D provides an estimated increment for the case where eye movements were 3D with the intercept being the reference category (i.e., 2D). Values in square brackets are 95\% CI.}}
\end{tabular}
\label{tab:regression}
\end{center}
\end{table}

\begin{figure}
    \centering
    \includegraphics[width=0.6\linewidth]{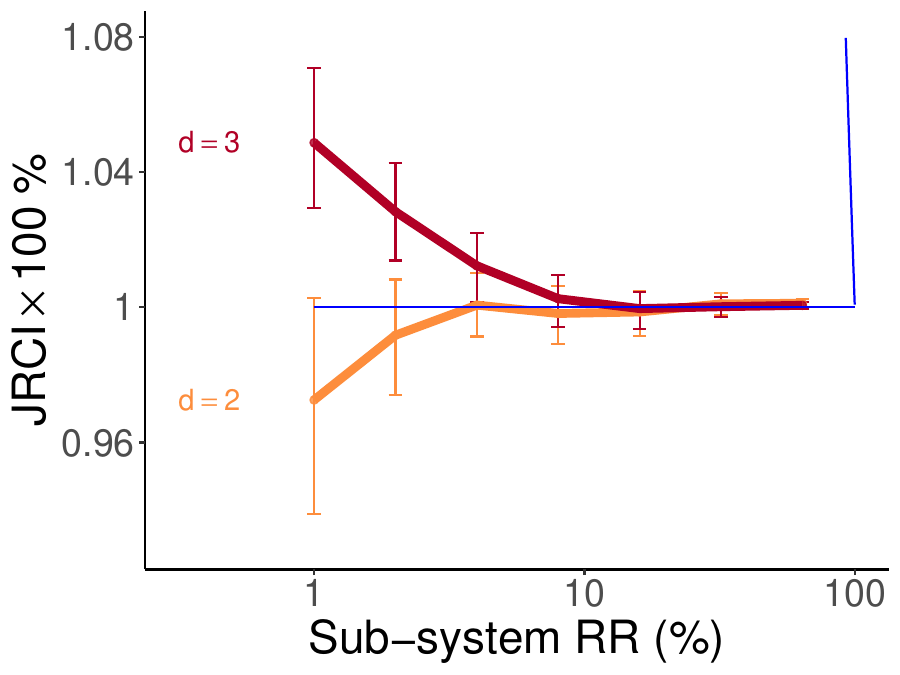}
    \caption{\textbf{Plot of JRCI vs.\ RR.} The graph shows that the coupling between EEG and 2D ($d = 2$) eye movements is compatible with or slightly below the random null model (horizontal blue line), whereas the coupling between EEG and 3D ($d = 3$) eye movements is above the random null model, albeit with quite weak coupling. Error bars are 95\% bootstrapped confidence intervals. Observations where subsystem $RR$ deviates from the specified value by more than 1.5 percentage points were excluded.}
    \label{fig:empirical-jrci}
\end{figure}

We estimated the model for subsystem $RR = 2\%$ and excluded observations where it was not possible to fix the recurrence rate with a tolerance of $\pm 1.5$ percentage points [see \cite{sm_methods} for details]. This led to the removal of 169 observations of EEG and 2D eye tracking, resulting in a total of 443 observations included in the model estimation. The results, shown in Table~\ref{tab:regression} and figure~\ref{fig:regression}, show a difference between 2D and 3D eye movements in relation to EEG: JRCI is higher for 3D eye movements and EEG compared to 2D eye movements and EEG, suggesting more systematic coupling of changes in 3D eye movements to changes in EEG. Because it was not always possible to fix subsystem recurrence at $2\%$ we performed additional robustness checks on the results to ensure that the conclusions do rely on the particulars of which observations were excluded (see table~\ref{tab:robustness}). The robustness checks give similar results to those presented in Table~\ref{tab:regression} [see \cite{sm_methods} for details]. While the regression analysis was performed for the particular choice of subsystem recurrence rate, $RR = 2\%$, other values give similar results, which is also seen in Figure~\ref{fig:empirical-jrci}, showing a plot of JRCI based on MvJRQA of EEG with 2D eye tracking and 3D eyetracking, respectively.

Of course, this is a proof-of-concept analysis, illustrating how MvJRQA can be used to capture coupling between different sets of multivariate time series differing in their dimensionality. Using these data, more fine-grained analysis might be performed, for example selecting sets of electrodes of the EEG that are primarily expected to relate to eye movements (e.g., occipital ones) and sets of electrodes that are primarily active in the planning processes that guide doctors' decision or assessment of a situation (e.g., frontal ones), and build a sequence of coupling analysis where frontal activity is strongly coupled to occipital activity, which in turn is strongly coupled to eye movements. These more advanced analyses are beyond the scope of the present work, but the current results show how such analyses can be conducted using MvJRQA to relate empirical time series of different dimensionality.

\section*{Discussion and Conclusion}
We introduced multivariate Joint Recurrence Quantification Analysis (MvJRQA), a method for correlating multivariate time series of different dimensionality. Based on four model systems, we showed that MvJRQA recovers coupling at the system level (i.e., when using all of the available observables) and that this works for both nonlinear and linear stochastic systems. Moreover, MvJRQA provides much more consistent results compared to simple Multidimensional Recurrence Quantification Analysis (MdRQA). We have also shown, in detail, how MvJRQA can be used for an empirical dataset where there is a large difference in the number of variables of the two systems and demonstrated a principled approach to using JRCI in a regression framework.

In applying the method it is important that the two RPs that are joined have an equal---or roughly equal---rate of recurrence. Otherwise, the RP with fewer recurrences will dictate the maximum number of possible joint recurrences, which will make the interpretation of the results more complicated. This goes particularly for data from larger samples that contain many instances of multivariate measurements.

The Joint Recurrence Coupling Indicator (JRCI) introduced in this paper generally helps to assess coupling by relating Joint Recurrence Rate ($JRR$) to subsystem recurrence of the individual systems, and thus allows to detect not only presence or absence, but also strength of coupling. Additionally, it ameliorates potential problems when subsystem $RR$ cannot be fully controlled, such as for categorical data, because the number of recurrences are---usually---a direct function of the data \cite{dale_nominal_2011}. In general, our simulations suggest that the JRCI works best when subsystem $RR$ is low, but not too low (i.e., between 1--5\%). However, our simulations do not cover all possible types of data, of course, so optimal values for subsystem $RR$ might be different for very different types of data or systems.

In general, the method works best under conditions where one is able to vary the coupling, or to detect a naturally occurring variation in the coupling between two systems. In our model systems, we could easily do that simply by changing the coupling constants in the models, but this is not necessarily possible with with empirical data. Instead it may be possible to exert control over some properties of the systems or their interaction, and MvJRQA can then be used to detect whether such as controlled change leads to a change in the joint recurrence rate, as we have shown for empirical data, the relationship between eye movements in two vs. three dimensions in relation to EEG records. 

In cases where it is impossible to apply exogenous experimental control it may instead be possible to use natural variation of the phenomenon being studied. In this case there should be some natural variation in the way the systems interact, that can be observed. Then it will be possible to perform an MvJRQA analysis of the combined system by breaking the time series down into epochs with different observed interactions---or alternatively using a continuously sliding window over the data for a windowed analysis. If one has only a single set of data, coupling can still be determined by the use of surrogate analysis, where coupling in empirical data is quantified against a baseline-model (see chapter 7 in \cite{wallot_recurrence-based_2025}) or against the random null model \cite{sm_methods}.

Note that relative recurrence loses the ability to detect or differentiate between coupling strengths in the case of extreme coupling, as is the case for many other methods that aim at detecting coupling \cite{monster_causal_2017}. 

\subsection*{Future research}
Having established the viability of applying MvJRQA to detect coupling between systems of arbitrary dimensions, we plan to extend the method to include other features of the joint recurrence plot than just the recurrence rate on which JRCI is based. For example the entropy of the distribution of lengths of diagonal lines may be informative of the complexity matching between the two coupled systems. For systems with time-varying coupling or time-lagged coupling we expect windowed analyses and time lagged analyses to provide a better representation of the interaction. In the present paper we have not addressed the issue of detecting coupling direction, which is obviously very important in many applications. To address this, extensions that incorporate the probability of the recurrence of one system conditional on recurrence of the other system will likely prove useful \cite{stam_synchronization_2002, romano_estimation_2007}. We have only briefly looked into the effect of observational noise, but for many applications dynamical noise is also important, since it can affect the systems' variables and propagate from one system to another. Therefore, assessing the effect of dynamical noise on the inferences obtained using MvJRQA is another important topic for future research. Since we are already able to correctly order pairs of sufficiently similar systems by relative coupling, a promising future development would be to test whether MvJRQA can successfully reconstruct a network model of system couplings based purely on observed time series. This would go beyond the current work, where we know (or assume) the delineation of the two sub systems a priori, as opposed to recovering system structure from the data. We have also focused on time series of continuous variables, but in some areas categorical time series may be required. One example is research where the behavior of participants is captured as a stream of different discrete actions that might then be correlated with physiological measures such as EEG. In order to use MvJRQA in such a case the method would need to be extended to handle categorical variables and a mixture of categorical and continuous variables. Finally, the algorithm can be improved to allow longer time series, to execute faster and find the phase space radius that results in a particular fixed recurrence rate with higher fidelity.

\subsection*{Materials and Methods}
In the following sections we provide information on the four models we used to simulate data, as well as details about the empirical dataset.

\subsubsection*{Model A: Two-dimensional and a one dimensional random process.}
To determine how well MvJRQA is able to detect correlations between a two-dimensional and a one-dimensional random process, we conducted a simple simulation where one system, reflected by one random variable, $x_1$ is composed of a single source of variation, $\varepsilon_1$. The other system is reflected by two random variables, $y_1$ and $y_2$. Each of these variables is composed of two sources of variation, one that provides idiosyncratic variability to each of the two variables, $\varepsilon_2$ and $\varepsilon_3$, respectively, and another random variable, $\varepsilon_4$, which provides a common source of variability to both variables, highlighting that $y_1$ and $y_2$ both belong to one system that introduces shared dynamics. To introduce coupling between the two systems, we use a weight $k$, which we varied between 0 (no correlation) and 10 (strong correlation) to change the correlation strength between the one-dimensional time series $x_1$ and the two two-dimensional time series $y_1$ and $y_2$. The variables are defined as follows:

\begin{align} \label{eq:linear}
    x_1 &= \varepsilon_1 \nonumber \\
    y_1 &= k x_1 + \varepsilon_2 + \varepsilon_4 \\
    y_2 &= k x_1 + \varepsilon_3 + \varepsilon_4 \nonumber
\end{align}
Here $k \in [0,10]$, and $\varepsilon_i$ are drawn from a Gaussian distribution with zero mean and unit variance:
\begin{equation}
\begin{array}{cc}
     \varepsilon_1 \sim N(0,1) \qquad& \varepsilon_2 \sim N(0,1)  \\
     \varepsilon_3 \sim N(0,1) \qquad& \varepsilon_4 \sim N(0,1)
\end{array}
\end{equation}

Because the data are random variables, no embedding is needed. Hence, the embedding dimension and the delay parameter are both set to one. The threshold parameter $\varepsilon_x$ for computing the uni-dimensional recurrence plot of $x_1$ was set to a value to yield approximately 10 percent recurrence for each plot, and the threshold parameter $\varepsilon_y$  for the multidimensional recurrence plot of $y_1$ and $y_2$ was likewise set to a value that yielded about the same percentage of recurrence points for each plot. This is done in order to give neither of the two RPs priority over the other (see section Multivariate Joint Recurrence Quantification Analysis (MvJRQA), above).

Using these settings, we varied the coupling parameter $k$ from 0 to 10 in step sizes of 0.5, and ran 100 instantiations of the random variables for each value of $k$. Each time series, $x_1$, $y_1$, and $y_2$, had a length of $N = 100$ data points.

In general, and as shown in the results, joint recurrence increases with increasing values of $k$. Moreover, we also notice that for no coupling (i.e., $k = 0$), we do not get 0 percent recurrence, but rather about 1 percent of recurrence. This is expected given the settings in our simulation: If we have two individual RPs whose computation is based on independent samples of stochastic data (in our case: $k = 0$), and each of these individual RPs yields at about 10 percent of recurrent point, then we expect their joint recurrence plot to yield 1 percent of recurrence. Accordingly, if we know the base recurrence rate of the two individual RPs ($RR_1$ and $RR_2$) to be joint, we can calculate the joint recurrence rate that we expect by chance ($JRR_\text{chance}$) simply by $JRR_\text{chance} = RR_1 \cdot RR_2$.

\subsubsection*{Model B: Periodic signal driving coupled logistic maps.}
The time evolution of dynamical systems can be modeled in either continuous time---flows, described by differential equations---or discrete time---maps, described by difference equations. In order to test how well MvJRQA is able to detect coupling in a discrete time system, we consider the logistic map\cite{may_simple_1976} which describes population growth with an optimal carrying capacity. In normalized units a the logistic map is given by the difference equation:
\begin{equation}
     x_{t+1} = r_x x_t (1-x_t),
\end{equation}
where $x_t \in [0, 1]$ is the population at time $t$ and $r_x \in [0, 4]$ is the growth rate. For low values of $r_x$ the logistic map converges to a fixed value and at intermediate values of $r_x$ stable periodic behavior with a period, always a power of 2, depending on $r_x$ is seen. At around $r \approx 3.56995$ the logistic map starts displaying chaotic behavior punctuated by periodic windows in $r_x$, including odd periods \cite{strogatz_nonlinear_1994}.  

We combine two logistic maps \cite{sugihara_detecting_2012} and add an external driving signal $H_t$ that modifies the growth rate:
\begin{equation}\label{eq:external}
\begin{split}
    x_{t+1} = x_t\left[ (r_x + \eta_x H_t)(1-x_t)-\beta_{xy}y_t\right]  \\
    y_{t+1} = y_t\left[ (r_y + \eta_y H_t)(1-y_t)-\beta_{yx}x_t\right]
\end{split}
\end{equation}
For simplicity, we have chosen the same coupling to the $x$ and $y$ variable, i.e., $\eta_x = \eta_y = \eta$. We use a previously studied external driver  \cite{summers_chaos_2000, monster_causal_2017}, viz. a cosine:
\begin{equation}\label{eq:moran-function}
H_t = \cos (2\pi t/p + \phi_0),
\end{equation}
where we have set the period to $p = 30$ and $\phi_0$ is a random initial phase. The endogenous growth rates  of the two logistic maps in Equation~\ref{eq:external} are kept constant at $r_x = 3.65$ and $r_y = 3.8$; and the couplings between them are fixed at $\beta_{xy} = 0$ and $\beta_{yx} = 0.4$, corresponding to unidirectional coupling from $x$ to $y$.

With this model we varied the coupling $\eta$ from the external driver to the two logistic maps. For each value of $\eta$ we simulated 100 systems with random initial conditions and sampled 500 values after discarding the first 300 samples to avoid transient dynamics related to the random initial conditions $(x_0, y_0, \phi_0)$. For MvJRQA the coupled logistic maps will be embedded in a 4-dimensional phase space and the external driver will be embedded into a 2-dimensional phase space. In both cases the time delay for the embedding is set to $\tau = 1$.

\subsubsection*{Model C: Lorenz system driving harmonic oscillator.}
The Lorenz system is described by coupled first-order differential equations for the three variables $x$, $y$, and $z$:
\begin{align} \label{eq:lorenz}
    \dot{x} &= \frac{\mathrm{d}x}{\mathrm{d}t} = \sigma \left( y - x \right) \nonumber \\
    \dot{y} &= \frac{\mathrm{d}y}{\mathrm{d}t} = x \left( \rho - z \right) - y \\
    \dot{z} &= \frac{\mathrm{d}z}{\mathrm{d}t} = xy - \beta z, \nonumber
\end{align}
where $\dot{x}$ is the derivative of $x$ with respect to time and $\sigma$, $\rho$, and $\beta$ are parameters in the model, which have the canonical values $\sigma = 10$, $\rho = 28$, and $\beta = 8/3$. The numerical solution to the three coupled differential equations in Equation~\ref{eq:lorenz} gives rise to the famous Lorenz butterfly attractor, shown as the phase space plot of system 1 in Figure~\ref{fig:conceptual}.

The harmonic oscillator is described by the second order differential equation 
\begin{equation}
    \Ddot{u} = \frac{\mathrm{d}^2u}{\mathrm{d}t^2} = -ku,
\end{equation}
for the variable $u$, where $k$ is a constant determining how big the force resulting from the displacement $u$ is. By introducing the velocity $v = \dot{u}$ the above second order differential equation can be written as two coupled first order differential equations, making explicit the two-dimensional nature of the harmonic oscillator system:
\begin{align}\label{eq:harmonic}
    \dot{u} &= v \qquad {}& \dot{v} &= -ku.
\end{align}

\begin{figure}
    \centering
    \includegraphics[width=\columnwidth]{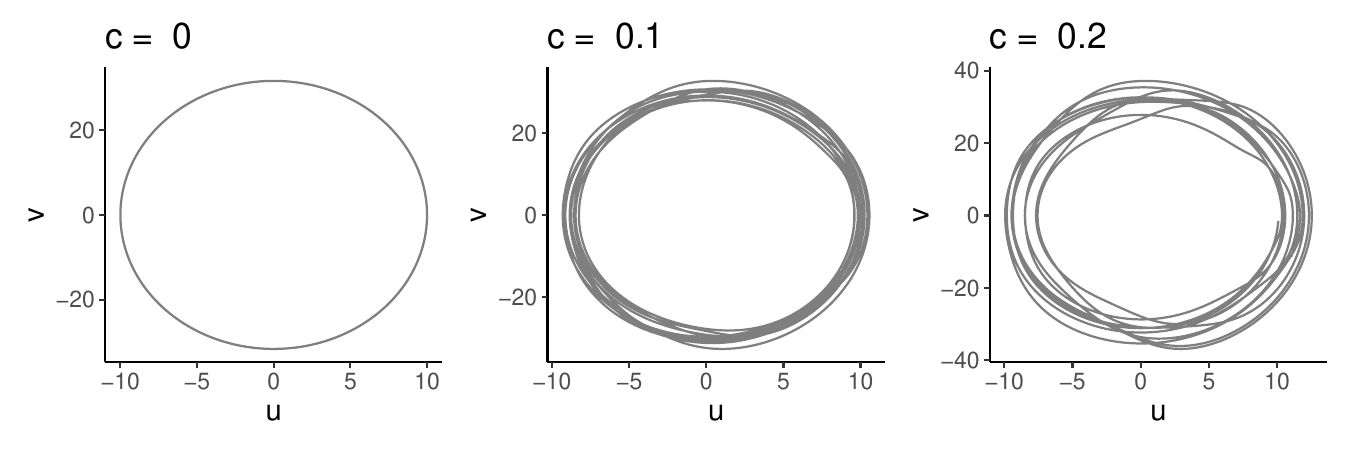}
    \caption{The phase space trajectory of the perturbed harmonic oscillator for different values of the coupling strength, $c$. For $c = 0$ the attractor is an ellipse, and with increasing values of $c$ the attractor becomes increasingly irregular as the harmonic oscillator is perturbed by the Lorenz system.}
    \label{fig:harmonic_ps}
\end{figure}

The combined model system is composed of the Lorenz system and harmonic oscillator system with a coupling term (force) proportional to $x^2$ that perturbs the harmonic oscillator:
\begin{align} \label{eq:lh_model}
    \dot{x} &= \sigma \left( y - x \right) \quad {}& \dot{u} &= v \nonumber \\
    \dot{y} &= x \left( \rho - z \right) - y \quad {}& \dot{v} &= -ku + cx^2 \\
    \dot{z} &= xy - \beta z {}& & \nonumber
\end{align}
These are the equations from Equation~\ref{eq:lorenz} and \ref{eq:harmonic} with the addition of the term $cx^2$ that is a unidirectional coupling from the Lorenz system to the harmonic oscillator system\footnote{For an example of a similar type of coupling, but to a different system, see \cite{marwan_nonlinear_2002}.}. The connections between the variables in the two coupled systems described by Equation~\ref{eq:lh_model} are illustrated in panel C of Figure~\ref{fig:model-timeseries}. We refer to $c$ as the coupling constant or coupling strength. In the limit of $c = 0$ we recover two separate, uncoupled, systems. The effect of increasing $c$ from zero to moderate ($c = 0.2$) to large ($c = 0.2$) values is shown in Figure~\ref{fig:harmonic_ps}, and in more detail in figure~\ref{fig:movie_snapshot} and movie~S1.

For each value of $c$ we simulated samples of length 500 for 100 different randomized initial conditions, while discarding the first 100 points. Both the Lorenz system and the harmonic oscillator were embedded in their own phase-space, as illustrated in Figure~\ref{fig:conceptual} using MdRQA to obtain their individual RPs.

\subsubsection*{Model D: Lorenz 96 system driving harmonic oscillator.}
The Lorenz 96 model \cite{lorenzPredictability2006} is defined as a set of $K$ nonlinear differential equations involving the variables $x_1, x_2, \ldots, x_K$:
\begin{equation} \label{eq:lorenz96}
    \dot{x}_k = \frac{\mathrm{d}x_k}{\mathrm{d}t} = (x_{k+1}-x_{k-2})x_{k-1} - x_k + F
\end{equation}
where $k \in [1, K]$ and $k$ is extended to all integer values by applying periodic boundary conditions, i.e., $x_{k + K} = x_{k - K} = x_k$.

The parameter $F$ is called the forcing, and we set it to $F = 8$, which gives rise to chaotic behavior. The model was introduced in 1996 by Edward Lorenz (hence the name) as a one-dimensional atmospheric model, where some quantity, $x$, is modeled in $K$ sectors of latitude. We include it here as a potentially high-dimensional nonlinear model, and include results for $K = 5$ and $K = 16$. We use the $x_1$-component of the Lorenz 96 system to drive a harmonic oscillator, in a manner similar to model system C:
\begin{equation} \label{eq:osc_l96}
    \begin{array}{rl}
         \dot{u} &= v \\
    \dot{v} &= -ku + \kappa x_1^2
    \end{array}
\end{equation}

As for the other systems, we vary the coupling, $\kappa$, systematically, and simulate 100 systems with random initial conditions, skipping the first 100 samples and recording 500 samples. The Lorenz 96 system is embedded as-is in a $K$-dimensional phase space and the harmonic oscillator in a 2-dimensional phase space.

\subsubsection*{Empirical data: Eye movements and EEG}
The dataset consists of co-registered EEG, as well as 2D and 3D eye movements for the 25 participants of the original study \cite{shafiei_performance_2023} who performed a variety of robotic surgery practice tasks. There is a variable number of trials per participant, as more junior participants (e.g., pre-medical students) performed the training task more often compared to the more experienced participants (e.g., faculty-level medical doctors) \cite{shafiei_performance_2023}. Data for participant 1 could not be used, as there were formatting issues with the EEG data. An additional seven trials were dropped because the data structures required for the analysis exceeded the \textsc{R} language's ability to index the large number of elements or because vectors were too long to pass to a \textsc{Fortran} subroutine. After this there were 612 person-trials (306 for each combination of trials of 2D and 3D eye movement coordinates with EEG). However, a further issue was lack of convergence, where subsystem recurrence could not be fixed within 1.5 \%-points of the desired recurrence rate of 2\% for 169 2D eye tracking observations, leaving a total of 443 trials for the main analysis. We explore this issue in depth in the Supplementary Materials \cite{sm_methods}, and provide robustness tests on the results of the regression analysis.

Starting with the raw data, eye movements were recorded at 50~Hz and EEG at 500~Hz. First, we trimmed the records and downsampled EEG to 50~Hz in order to obtain matched time series with the same number of data points for each trial. Then, all occulogram data were removed from the 124-channel EEG records (see figure~\ref{fig:empirical-timeseries} for example time series). All data, eye movement coordinates and EEG, were differenced, as it does not seem reasonable to assume coupling of specific locations (coordinates) of the eye movement data with specific micro-voltage levels in the EEG, but rather that transitions in eye movements are linked to transitions in the brain data. Finally, missing values were removed in the differenced records.

Next, data was subjected to MvJRQA with 2D eye movements and EEG, as well as 3D eye movements and EEG for each participant and trial. For the MvJRQA analysis, we fixed the individual recurrence percentage at 2\% for each individual multivariate time series, i.e., 2D eye movements, 3D eye movements and EEG [see \cite{sm_methods} for details]. This did not work in all cases, particularly because of the nature of the eye movement data, where changing periods of fixations and saccades lead to signal dynamics with discontinuous increases of recurrence given a slight increase of the radius parameter. On average, percentage recurrence was 2.03\% ($\text{SD} = 0.27$) for the observations used in the main regression analyses reported in Table~\ref{tab:regression}. To conduct MvJRQA, we set the the delay parameter and embedding dimension parameter both to 1 and the data were not z-scored before analysis, as the eye movement coordinates and EEG data are naturally scaled to one another.

\clearpage %

\bibliography{references.bib, sm.bib}

\begin{thebibliography}{10}
\providecommand{\url}[1]{\texttt{#1}}
\expandafter\ifx\csname urlstyle\endcsname\relax
  \providecommand{\doi}[1]{doi:\discretionary{}{}{}#1}\else
  \providecommand{\doi}{doi:\discretionary{}{}{}\begingroup
  \urlstyle{rm}\Url}\fi

\bibitem{marwan_recurrence_2007}
N.~Marwan, M.~Carmen~Romano, M.~Thiel, J.~Kurths, Recurrence plots for the
  analysis of complex systems. \emph{Physics Reports} \textbf{438}~(5),
  237--329 (2007), \doi{10.1016/j.physrep.2006.11.001}.

\bibitem{wallot_multidimensional_2016}
S.~Wallot, A.~Roepstorff, D.~Mønster, Multidimensional {Recurrence}
  {Quantification} {Analysis} ({MdRQA}) for the {Analysis} of
  {Multidimensional} {Time}-{Series}: {A} {Software} {Implementation} in
  {MATLAB} and {Its} {Application} to {Group}-{Level} {Data} in {Joint}
  {Action}. \emph{Frontiers in Psychology} \textbf{7} (2016),
  \doi{10.3389/fpsyg.2016.01835},
  \url{https://www.frontiersin.org/article/10.3389/fpsyg.2016.01835}.

\bibitem{mayseless_real-life_2019}
N.~Mayseless, G.~Hawthorne, A.~L. Reiss, Real-life creative problem solving in
  teams: {fNIRS} based hyperscanning study. \emph{NeuroImage} \textbf{203},
  116161 (2019), \doi{10.1016/j.neuroimage.2019.116161}.

\bibitem{fuchs_coordination_2018}
A.~Fuchs, J.~A. Scott~Kelso, Coordination {Dynamics} and {Synergetics}: {From}
  {Finger} {Movements} to {Brain} {Patterns} and {Ballet} {Dancing}, in
  \emph{Complexity and {Synergetics}}, S.~C. Müller, P.~J. Plath, G.~Radons,
  A.~Fuchs, Eds. (Springer International Publishing, Cham), pp. 301--316
  (2018), \doi{10.1007/978-3-319-64334-2_23},
  \url{https://doi.org/10.1007/978-3-319-64334-2_23}.

\bibitem{scholz_intentional_1990}
J.~P. Scholz, J.~A.~S. Kelso, Intentional {Switching} {Between} {Patterns} of
  {Bimanual} {Coordination} {Depends} on the {Intrinsic} {Dynamics} of the
  {Patterns}. \emph{Journal of Motor Behavior} \textbf{22}~(1), 98--124 (1990),
  \doi{10.1080/00222895.1990.10735504}.

\bibitem{ramenzoni_joint_2011}
V.~C. Ramenzoni, T.~J. Davis, M.~A. Riley, K.~Shockley, A.~A. Baker, Joint
  action in a cooperative precision task: nested processes of intrapersonal and
  interpersonal coordination. \emph{Experimental Brain Research}
  \textbf{211}~(3), 447--457 (2011), \doi{10.1007/s00221-011-2653-8}.

\bibitem{crone_synchronous_2021}
C.~L. Crone, \emph{et~al.}, Synchronous vs. non-synchronous imitation: {Using}
  dance to explore interpersonal coordination during observational learning.
  \emph{Human Movement Science} \textbf{76}, 102776 (2021),
  \doi{10.1016/j.humov.2021.102776},
  \url{https://www.sciencedirect.com/science/article/pii/S0167945721000245}.

\bibitem{stephen_dynamics_2009}
D.~G. Stephen, R.~A. Boncoddo, J.~S. Magnuson, J.~A. Dixon, The dynamics of
  insight: {Mathematical} discovery as a phase transition. \emph{Memory \&
  Cognition} \textbf{37}~(8), 1132--1149 (2009), \doi{10.3758/MC.37.8.1132},
  \url{https://doi.org/10.3758/MC.37.8.1132}.

\bibitem{kuznetsov_effects_2011}
N.~Kuznetsov, S.~Wallot, Effects of {Accuracy} {Feedback} on {Fractal}
  {Characteristics} of {Time} {Estimation}. \emph{Frontiers in Integrative
  Neuroscience} \textbf{5} (2011), \doi{10.3389/fnint.2011.00062}.

\bibitem{kello_pervasiveness_2008}
C.~T. Kello, G.~G. Anderson, J.~G. Holden, G.~C. Van~Orden, The {Pervasiveness}
  of 1/f {Scaling} in {Speech} {Reflects} the {Metastable} {Basis} of
  {Cognition}. \emph{Cognitive Science} \textbf{32}~(7), 1217--1231 (2008),
  \doi{10.1080/03640210801944898}.

\bibitem{he_scale-free_2011}
B.~J. He, Scale-{Free} {Properties} of the {Functional} {Magnetic} {Resonance}
  {Imaging} {Signal} during {Rest} and {Task}. \emph{Journal of Neuroscience}
  \textbf{31}~(39), 13786--13795 (2011), \doi{10.1523/JNEUROSCI.2111-11.2011},
  \url{https://www.jneurosci.org/content/31/39/13786}.

\bibitem{abney_complexity_2014}
D.~H. Abney, A.~Paxton, R.~Dale, C.~T. Kello, Complexity matching in dyadic
  conversation. \emph{Journal of Experimental Psychology: General}
  \textbf{143}, 2304--2315 (2014), \doi{10.1037/xge0000021}.

\bibitem{marmelat_strong_2012}
V.~Marmelat, D.~Delignières, Strong anticipation: complexity matching in
  interpersonal coordination. \emph{Experimental Brain Research}
  \textbf{222}~(1), 137--148 (2012), \doi{10.1007/s00221-012-3202-9}.

\bibitem{delignieres_multifractal_2016}
D.~Delignières, Z.~M.~H. Almurad, C.~Roume, V.~Marmelat, Multifractal
  signatures of complexity matching. \emph{Experimental Brain Research}
  \textbf{234}~(10), 2773--2785 (2016), \doi{10.1007/s00221-016-4679-4},
  \url{https://doi.org/10.1007/s00221-016-4679-4}.

\bibitem{ihlen_interaction-dominant_2010}
E.~A.~F. Ihlen, B.~Vereijken, Interaction-dominant dynamics in human cognition:
  {Beyond} 1/f fluctuation. \emph{Journal of Experimental Psychology: General}
  \textbf{139}, 436--463 (2010), \doi{10.1037/a0019098}.

\bibitem{nelson-wong_application_2009}
E.~Nelson-Wong, S.~Howarth, D.~A. Winter, J.~P. Callaghan, Application of
  {Autocorrelation} and {Cross}-correlation {Analyses} in {Human} {Movement}
  and {Rehabilitation} {Research}. \emph{Journal of Orthopaedic \& Sports
  Physical Therapy} \textbf{39}~(4), 287--295 (2009),
  \doi{10.2519/jospt.2009.2969},
  \url{https://www.jospt.org/doi/abs/10.2519/jospt.2009.2969}.

\bibitem{granger_investigating_1969}
C.~W.~J. Granger, Investigating {Causal} {Relations} by {Econometric} {Models}
  and {Cross}-spectral {Methods}. \emph{Econometrica} \textbf{37}~(3), 424--438
  (1969), publisher: [Wiley, Econometric Society], \doi{10.2307/1912791},
  \url{https://www.jstor.org/stable/1912791}.

\bibitem{schreiber_measuring_2000}
T.~Schreiber, Measuring {Information} {Transfer}. \emph{Physical Review
  Letters} \textbf{85}~(2), 461--464 (2000), publisher: American Physical
  Society, \doi{10.1103/PhysRevLett.85.461},
  \url{https://link.aps.org/doi/10.1103/PhysRevLett.85.461}.

\bibitem{frenzel_partial_2007}
S.~Frenzel, B.~Pompe, Partial {Mutual} {Information} for {Coupling} {Analysis}
  of {Multivariate} {Time} {Series}. \emph{Physical Review Letters}
  \textbf{99}~(20), 204101 (2007), publisher: American Physical Society,
  \doi{10.1103/PhysRevLett.99.204101},
  \url{https://link.aps.org/doi/10.1103/PhysRevLett.99.204101}.

\bibitem{shockley_cross_2002}
K.~Shockley, M.~Butwill, J.~P. Zbilut, C.~L. Webber, Cross recurrence
  quantification of coupled oscillators. \emph{Physics Letters A}
  \textbf{305}~(1), 59--69 (2002), \doi{10.1016/S0375-9601(02)01411-1},
  \url{https://www.sciencedirect.com/science/article/pii/S0375960102014111}.

\bibitem{sugihara_detecting_2012}
G.~Sugihara, \emph{et~al.}, Detecting {Causality} in {Complex} {Ecosystems}.
  \emph{Science} \textbf{338}~(6106), 496--500 (2012),
  \doi{10.1126/science.1227079},
  \url{https://www.science.org/doi/full/10.1126/science.1227079}.

\bibitem{paxton_network_2014}
A.~Paxton, D.~H. Abney, C.~T. Kello, R.~K. Dale, Network {Analysis} of
  {Multimodal}, {Multiscale} {Coordination} in {Dyadic} {Problem} {Solving}.
  \emph{Proceedings of the Annual Meeting of the Cognitive Science Society}
  \textbf{36}~(36) (2014), \url{https://escholarship.org/uc/item/7xz2z06w}.

\bibitem{vanhatalo_impact_2016}
E.~Vanhatalo, M.~Kulahci, Impact of {Autocorrelation} on {Principal}
  {Components} and {Their} {Use} in {Statistical} {Process} {Control}.
  \emph{Quality and Reliability Engineering International} \textbf{32}~(4),
  1483--1500 (2016), \doi{10.1002/qre.1858},
  \url{https://onlinelibrary.wiley.com/doi/abs/10.1002/qre.1858}.

\bibitem{thompson_canonical_1984}
B.~Thompson, \emph{Canonical {Correlation} {Analysis}: {Uses} and
  {Interpretation}} (SAGE) (1984).

\bibitem{wallot_multidimensional_2019}
S.~Wallot, Multidimensional {Cross}-{Recurrence} {Quantification} {Analysis}
  ({MdCRQA}) – {A} {Method} for {Quantifying} {Correlation} between
  {Multivariate} {Time}-{Series}. \emph{Multivariate Behavioral Research}
  \textbf{54}~(2), 173--191 (2019), \doi{10.1080/00273171.2018.1512846},
  \url{https://doi.org/10.1080/00273171.2018.1512846}.

\bibitem{wallot_using_2013}
S.~Wallot, R.~Fusaroli, K.~Tylen, E.-M. Jegindø, Using complexity metrics with
  {R}-{R} intervals and {BPM} heart rate measures. \emph{Frontiers in
  Physiology} \textbf{4} (2013), \doi{doi.org/10.3389/fphys.2013.00211},
  \url{https://www.frontiersin.org/article/10.3389/fphys.2013.00211}.

\bibitem{stam_synchronization_2002}
C.~J. Stam, B.~W. van Dijk, Synchronization likelihood: an unbiased measure of
  generalized synchronization in multivariate data sets. \emph{Physica D:
  Nonlinear Phenomena} \textbf{163}~(3), 236--251 (2002),
  \doi{10.1016/S0167-2789(01)00386-4},
  \url{https://www.sciencedirect.com/science/article/pii/S0167278901003864}.

\bibitem{romano_detection_2005}
M.~C. Romano, M.~Thiel, J.~Kurths, I.~Z. Kiss, J.~L. Hudson, Detection of
  synchronization for non-phase-coherent and non-stationary data.
  \emph{Europhysics Letters} \textbf{71}~(3), 466 (2005), publisher: IOP
  Publishing, \doi{10.1209/epl/i2005-10095-1},
  \url{https://iopscience.iop.org/article/10.1209/epl/i2005-10095-1/meta}.

\bibitem{romano_estimation_2007}
M.~C. Romano, M.~Thiel, J.~Kurths, C.~Grebogi, Estimation of the direction of
  the coupling by conditional probabilities of recurrence. \emph{Physical
  Review E} \textbf{76}~(3), 036211 (2007), publisher: American Physical
  Society, \doi{10.1103/PhysRevE.76.036211},
  \url{https://link.aps.org/doi/10.1103/PhysRevE.76.036211}.

\bibitem{runge_detecting_2019}
J.~Runge, P.~Nowack, M.~Kretschmer, S.~Flaxman, D.~Sejdinovic, Detecting and
  quantifying causal associations in large nonlinear time series datasets.
  \emph{Science Advances} \textbf{5}~(11), eaau4996 (2019), publisher: American
  Association for the Advancement of Science, \doi{10.1126/sciadv.aau4996},
  \url{https://www.science.org/doi/full/10.1126/sciadv.aau4996}.

\bibitem{gilpin_recurrences_2025}
W.~Gilpin, Recurrences {Reveal} {Shared} {Causal} {Drivers} of {Complex} {Time}
  {Series}. \emph{Physical Review X} \textbf{15}~(1), 011005 (2025), publisher:
  American Physical Society, \doi{10.1103/PhysRevX.15.011005},
  \url{https://link.aps.org/doi/10.1103/PhysRevX.15.011005}.

\bibitem{romano_multivariate_2004}
M.~C. Romano, M.~Thiel, J.~Kurths, W.~von Bloh, Multivariate recurrence plots.
  \emph{Physics Letters A} \textbf{330}~(3), 214--223 (2004),
  \doi{10.1016/j.physleta.2004.07.066}.

\bibitem{takens_detecting_1981}
F.~Takens, Detecting strange attractors in turbulence. \emph{Lecture Notes in
  Mathematics} \textbf{898}, 366--381 (1981), \doi{10.1007/BFb0091924}.

\bibitem{wallot_calculation_2018}
S.~Wallot, D.~Mønster, Calculation of {Average} {Mutual} {Information} ({AMI})
  and {False}-{Nearest} {Neighbors} ({FNN}) for the {Estimation} of {Embedding}
  {Parameters} of {Multidimensional} {Time} {Series} in {Matlab}.
  \emph{Frontiers in Psychology} \textbf{9} (2018),
  \url{https://www.frontiersin.org/article/10.3389/fpsyg.2018.01679}.

\bibitem{eckmann_recurrence_1987}
J.-P. Eckmann, S.~O. Kamphorst, D.~Ruelle, Recurrence {Plots} of {Dynamical}
  {Systems}. \emph{Europhysics Letters (EPL)} \textbf{4}~(9), 973--977 (1987),
  \doi{10.1209/0295-5075/4/9/004},
  \url{https://doi.org/10.1209/0295-5075/4/9/004}.

\bibitem{webber_dynamical_1994}
C.~L. Webber, J.~P. Zbilut, Dynamical assessment of physiological systems and
  states using recurrence plot strategies. \emph{Journal of Applied Physiology}
  \textbf{76}~(2), 965--973 (1994), \doi{10.1152/jappl.1994.76.2.965},
  \url{https://journals.physiology.org/doi/abs/10.1152/jappl.1994.76.2.965}.

\bibitem{zbilut_detecting_1998}
J.~P. Zbilut, A.~Giuliani, C.~L. Webber, Detecting deterministic signals in
  exceptionally noisy environments using cross-recurrence quantification.
  \emph{Physics Letters A} \textbf{246}~(1), 122--128 (1998),
  \doi{10.1016/S0375-9601(98)00457-5},
  \url{https://www.sciencedirect.com/science/article/pii/S0375960198004575}.

\bibitem{coco_r_2021}
M.~I. Coco, D.~Mønster, G.~Leonardi, R.~Dale, S.~Wallot, The {R} {Journal}:
  {Unidimensional} and {Multidimensional} {Methods} for {Recurrence}
  {Quantification} {Analysis} with crqa. \emph{The R Journal} \textbf{13}~(1),
  145--163 (2021), \doi{10.32614/RJ-2021-062},
  \url{https://doi.org/10.32614/RJ-2021-062/}.

\bibitem{monster_mvjrqa_2025}
D.~Mønster, S.~Wallot, {MvJRQA} (2025), \doi{10.5281/zenodo.17724340},
  \url{https://zenodo.org/records/17724340}.

\bibitem{mcardle_latent_2009}
J.~J. McArdle, Latent {Variable} {Modeling} of {Differences} and {Changes} with
  {Longitudinal} {Data}. \emph{Annual Review of Psychology} \textbf{60}~(1),
  577--605 (2009), \doi{10.1146/annurev.psych.60.110707.163612},
  \url{https://doi.org/10.1146/annurev.psych.60.110707.163612}.

\bibitem{may_simple_1976}
R.~M. May, Simple mathematical models with very complicated dynamics.
  \emph{Nature} \textbf{261}~(5560), 459--467 (1976), publisher: Nature
  Publishing Group, \doi{10.1038/261459a0},
  \url{https://www.nature.com/articles/261459a0}.

\bibitem{summers_chaos_2000}
D.~Summers, J.~G. Cranford, B.~P. Healey, Chaos in periodically forced
  discrete-time ecosystem models. \emph{Chaos, Solitons \& Fractals}
  \textbf{11}~(14), 2331--2342 (2000), \doi{10.1016/S0960-0779(99)00154-X},
  \url{https://www.sciencedirect.com/science/article/pii/S096007799900154X}.

\bibitem{monster_causal_2017}
D.~Mønster, R.~Fusaroli, K.~Tylén, A.~Roepstorff, J.~F. Sherson, Causal
  inference from noisy time-series data — {Testing} the {Convergent}
  {Cross}-{Mapping} algorithm in the presence of noise and external influence.
  \emph{Future Generation Computer Systems} \textbf{73}, 52--62 (2017),
  \doi{10.1016/j.future.2016.12.009}.

\bibitem{lorenz_deterministic_1963}
E.~N. Lorenz, Deterministic {Nonperiodic} {Flow}. \emph{Journal of the
  Atmospheric Sciences} \textbf{20}~(2), 130--141 (1963),
  \doi{10.1175/1520-0469(1963)020<0130:DNF>2.0.CO;2},
  \url{https://journals.ametsoc.org/view/journals/atsc/20/2/1520-0469_1963_020_0130_dnf_2_0_co_2.xml}.

\bibitem{lorenzPredictability2006}
E.~N. Lorenz, Predictability – a problem partly solved, in
  \emph{Predictability of {Weather} and {Climate}}, R.~Hagedorn, T.~Palmer,
  Eds. (Cambridge University Press, Cambridge), pp. 40--58 (2006),
  \doi{10.1017/CBO9780511617652.004},
  \url{https://www.cambridge.org/core/books/predictability-of-weather-and-climate/predictability-a-problem-partly-solved/3221BDE379DEB669BA52C66263AF3206}.

\bibitem{sm_methods}
Materials and methods are available as supplementary material.

\bibitem{bresar_directional_2023}
M.~Brešar, P.~Boškoski, Directional coupling detection through cross-distance
  vectors. \emph{Physical Review E} \textbf{107}~(4), 044220 (2023), publisher:
  American Physical Society, \doi{10.1103/PhysRevE.107.044220},
  \url{https://link.aps.org/doi/10.1103/PhysRevE.107.044220}.

\bibitem{shafiei_integration_2023}
S.~B. Shafiei, S.~Shadpour, Integration of {Electroencephalogram} and
  {Eye}-{Gaze} {Datasets} for {Performance} {Evaluation} in {Fundamentals} of
  {Laparoscopic} {Surgery} ({FLS}) {Tasks} (version 1.0.0) (2023),
  \doi{10.13026/C8SA-ED22},
  \url{https://physionet.org/content/eeg-eye-gaze-for-fls-tasks/}, physioNet.

\bibitem{goldberger_physiobank_2000}
A.~L. Goldberger, \emph{et~al.}, {PhysioBank}, {PhysioToolkit}, and
  {PhysioNet}. \emph{Circulation} \textbf{101}~(23), e215--e220 (2000),
  \doi{10.1161/01.CIR.101.23.e215},
  \url{https://www.ahajournals.org/doi/10.1161/01.CIR.101.23.e215}.

\bibitem{shafiei_performance_2023}
S.~B. Shafiei, \emph{et~al.}, Performance and learning rate prediction models
  development in {FLS} and {RAS} surgical tasks using electroencephalogram and
  eye gaze data and machine learning. \emph{Surgical Endoscopy}
  \textbf{37}~(11), 8447--8463 (2023), \doi{10.1007/s00464-023-10409-y},
  \url{https://doi.org/10.1007/s00464-023-10409-y}.

\bibitem{dale_nominal_2011}
R.~Dale, A.~S. Warlaumont, D.~C. Richardson, Nominal cross recurrence as a
  generalized lag sequential analysis for behavioral streams.
  \emph{International Journal of Bifurcation and Chaos} \textbf{21}~(04),
  1153--1161 (2011), \doi{10.1142/S0218127411028970},
  \url{https://www.worldscientific.com/doi/abs/10.1142/S0218127411028970}.

\bibitem{wallot_recurrence-based_2025}
S.~Wallot, G.~Leonardi, \emph{Recurrence-{Based} {Analyses}} (SAGE
  Publications, London) (2025).

\bibitem{strogatz_nonlinear_1994}
S.~H. Strogatz, \emph{Nonlinear dynamics and chaos: with applications to
  physics, biology, chemistry, and engineering}, Studies in nonlinearity
  (Westview Press, Cambridge (Mass.)) (1994).

\bibitem{marwan_nonlinear_2002}
N.~Marwan, J.~Kurths, Nonlinear analysis of bivariate data with cross
  recurrence plots. \emph{Physics Letters A} \textbf{302}~(5), 299--307 (2002),
  \doi{10.1016/S0375-9601(02)01170-2},
  \url{https://www.sciencedirect.com/science/article/pii/S0375960102011702}.

\bibitem{szpiro_measuring_1993}
G.~G. Szpiro, Measuring dynamical noise in dynamical systems. \emph{Physica D:
  Nonlinear Phenomena} \textbf{65}~(3), 289--299 (1993),
  \doi{10.1016/0167-2789(93)90164-V},
  \url{https://www.sciencedirect.com/science/article/pii/016727899390164V}.

\bibitem{r_core_team_r_nodate}
{R Core Team}, R: {The} {R} project for statistical computing,
  \url{https://www.r-project.org/}.

\bibitem{ushey_renv_2025}
K.~Ushey, H.~Wickham, renv: {Project} {Environments} (2025),
  \url{https://cran.r-project.org/web/packages/renv/index.html}.

\bibitem{xie_animation_2013}
Y.~Xie, animation: {An} {R} {Package} for {Creating} {Animations} and
  {Demonstrating} {Statistical} {Methods}. \emph{Journal of Statistical
  Software} \textbf{53}, 1--27 (2013), \doi{10.18637/jss.v053.i01},
  \url{https://doi.org/10.18637/jss.v053.i01}.

\bibitem{wilke_cowplot_2025}
C.~O. Wilke, cowplot: {Streamlined} {Plot} {Theme} and {Plot} {Annotations} for
  'ggplot2' (2025),
  \url{https://cran.r-project.org/web/packages/cowplot/index.html}.

\bibitem{barrett_datatable_2025}
T.~Barrett, \emph{et~al.}, data.table: {Extension} of 'data.frame' (2025),
  \url{https://cran.r-project.org/web/packages/data.table/index.html}.

\bibitem{soetaert_solving_2010}
K.~Soetaert, T.~Petzoldt, R.~W. Setzer, Solving {Differential} {Equations} in
  {R}: {Package} {deSolve}. \emph{Journal of Statistical Software} \textbf{33},
  1--25 (2010), \doi{10.18637/jss.v033.i09},
  \url{https://doi.org/10.18637/jss.v033.i09}.

\bibitem{wickham_dplyr_2023}
H.~Wickham, \emph{et~al.}, dplyr: {A} {Grammar} of {Data} {Manipulation}
  (2023), \url{https://cran.r-project.org/web/packages/dplyr/index.html}.

\bibitem{henelius_edf_2016}
A.~Henelius, edf: {Read} {Data} from {European} {Data} {Format} ({EDF} and
  {EDF}+) {Files} (2016),
  \url{https://cran.r-project.org/web/packages/edf/index.html}.

\bibitem{ben-shachar_effectsize_2020}
M.~S. Ben-Shachar, D.~Lüdecke, D.~Makowski, effectsize: {Estimation} of
  {Effect} {Size} {Indices} and {Standardized} {Parameters}. \emph{Journal of
  Open Source Software} \textbf{5}~(56), 2815 (2020),
  \doi{10.21105/joss.02815},
  \url{https://joss.theoj.org/papers/10.21105/joss.02815}.

\bibitem{auguie_gridextra_2017}
B.~Auguie, A.~Antonov, {gridExtra}: {Miscellaneous} {Functions} for "{Grid}"
  {Graphics} (2017),
  \url{https://cran.r-project.org/web/packages/gridExtra/index.html}.

\bibitem{wickham_ggplot2_2016}
H.~Wickham, \emph{ggplot2: {Elegant} {Graphics} for {Data} {Analysis}}
  (Springer) (2016).

\bibitem{slowikowski_ggrepel_2024}
K.~Slowikowski, \emph{et~al.}, ggrepel: {Automatically} {Position}
  {Non}-{Overlapping} {Text} {Labels} with 'ggplot2' (2024),
  \url{https://cran.r-project.org/web/packages/ggrepel/index.html}.

\bibitem{meschiari_latex2exp_2022}
S.~Meschiari, latex2exp: {Use} {LaTeX} {Expressions} in {Plots} (2022),
  \url{https://cran.r-project.org/web/packages/latex2exp/index.html}.

\bibitem{bates_fitting_2015}
D.~Bates, M.~Mächler, B.~Bolker, S.~Walker, Fitting {Linear} {Mixed}-{Effects}
  {Models} {Using} lme4. \emph{Journal of Statistical Software} \textbf{67},
  1--48 (2015), \doi{10.18637/jss.v067.i01},
  \url{https://doi.org/10.18637/jss.v067.i01}.

\bibitem{kuznetsova_lmertest_2017}
A.~Kuznetsova, P.~B. Brockhoff, R.~H.~B. Christensen, {lmerTest} {Package}:
  {Tests} in {Linear} {Mixed} {Effects} {Models}. \emph{Journal of Statistical
  Software} \textbf{82}, 1--26 (2017), \doi{10.18637/jss.v082.i13},
  \url{https://doi.org/10.18637/jss.v082.i13}.

\bibitem{bates_matrix_2025}
D.~Bates, \emph{et~al.}, Matrix: {Sparse} and {Dense} {Matrix} {Classes} and
  {Methods} (2025),
  \url{https://cran.r-project.org/web/packages/Matrix/index.html}.

\bibitem{makowski_modelbased_2025}
D.~Makowski, \emph{et~al.}, modelbased: {An} {R} package to make the most out
  of your statistical models through marginal means, marginal effects, and
  model predictions. \emph{Journal of Open Source Software} \textbf{10}~(109),
  7969 (2025), \doi{10.21105/joss.07969},
  \url{https://joss.theoj.org/papers/10.21105/joss.07969}.

\bibitem{pedersen_patchwork_2025}
T.~L. Pedersen, patchwork: {The} {Composer} of {Plots} (2025),
  \url{https://cran.r-project.org/web/packages/patchwork/index.html}.

\bibitem{ludecke_performance_2021}
D.~Lüdecke, M.~S. Ben-Shachar, I.~Patil, P.~Waggoner, D.~Makowski,
  performance: {An} {R} {Package} for {Assessment}, {Comparison} and {Testing}
  of {Statistical} {Models}. \emph{Journal of Open Source Software}
  \textbf{6}~(60), 3139 (2021), \doi{10.21105/joss.03139},
  \url{https://joss.theoj.org/papers/10.21105/joss.03139}.

\bibitem{wickham_purrr_2025}
H.~Wickham, L.~Henry, P.~Software, P.~[cph, fnd, purrr: {Functional}
  {Programming} {Tools} (2025),
  \url{https://cran.r-project.org/web/packages/purrr/index.html}.

\bibitem{wickham_readr_2024}
H.~Wickham, \emph{et~al.}, readr: {Read} {Rectangular} {Text} {Data} (2024),
  \url{https://cran.r-project.org/web/packages/readr/index.html}.

\bibitem{murdoch_rgl_2025}
D.~Murdoch, \emph{et~al.}, rgl: {3D} {Visualization} {Using} {OpenGL} (2025),
  \url{https://cran.r-project.org/web/packages/rgl/index.html}.

\bibitem{ludecke_see_2021}
D.~Lüdecke, \emph{et~al.}, see: {An} {R} {Package} for {Visualizing}
  {Statistical} {Models}. \emph{Journal of Open Source Software}
  \textbf{6}~(64), 3393 (2021), \doi{10.21105/joss.03393},
  \url{https://joss.theoj.org/papers/10.21105/joss.03393}.

\bibitem{leifeld_texreg_2013}
P.~Leifeld, texreg: {Conversion} of {Statistical} {Model} {Output} in {R} to
  {LATEX} and {HTML} {Tables}. \emph{Journal of Statistical Software}
  \textbf{55}, 1--24 (2013), \doi{10.18637/jss.v055.i08},
  \url{https://doi.org/10.18637/jss.v055.i08}.

\bibitem{wickham_tidyr_2024}
H.~Wickham, \emph{et~al.}, tidyr: {Tidy} {Messy} {Data} (2024),
  \url{https://cran.r-project.org/web/packages/tidyr/index.html}.

\end{thebibliography}
\bibliographystyle{sciencemag}

We wish to thank the organizers and attendees of the 10th International Recurrence Plot Symposium held at the University of Tsukuba, Japan, who provided useful feedback on an earlier version  of this research.

\paragraph*{Funding:}
SW acknowledges funding from the German Science Foundation (Deutsche Forschungsgemeinschaft, DFG), Heisenberg grant 442405852 and research grant project 442405919.

\paragraph*{Author contributions:}
SW and DM contributed equally to all aspects of this work.

\paragraph*{Competing interests:}
There are no competing interests to declare.

\paragraph*{Data and materials availability:}
Code and replication scripts are available on Zenodo \cite{monster_mvjrqa_2025} (\url{https://zenodo.org/records/17724340}) and GitHub (\url{https://github.com/danm0nster/mvjrqa-replication}). 
Empirical data are available from PhysioNet \cite{goldberger_physiobank_2000, shafiei_integration_2023} (\url{https://physionet.org/content/eeg-eye-gaze-for-fls-tasks/1.0.0/}).

\subsection*{Supplementary materials}
Materials and Methods\\
Supplementary Text\\
Figs. S1 to S8\\
Table S1\\
References \textit{(55-\arabic{enumiv})}\\ %
Movie S1%

\newpage

\renewcommand{\thefigure}{S\arabic{figure}}
\renewcommand{\thetable}{S\arabic{table}}
\renewcommand{\theequation}{S\arabic{equation}}
\renewcommand{\thepage}{S\arabic{page}}
\setcounter{figure}{0}
\setcounter{table}{0}
\setcounter{equation}{0}
\setcounter{page}{1} %

\begin{center}
\section*{Supplementary Materials for\\ \scititle}

Sebastian Wallot$^{\dagger}$,
Dan M{\o}nster$^{\ast\dagger}$\\ %
\small$^\ast$Corresponding author. Email: danm@econ.au.dk\\
\small$^\dagger$These authors contributed equally to this work.
\end{center}

\subsubsection*{This PDF file includes:}
Materials and Methods\\
Figures S1 to S8\\
Table S1\\
Caption for Movie S1\\

\subsubsection*{Other Supplementary Materials for this manuscript:}
Movie S1\\

\newpage

\subsection*{Materials and Methods}

\subsubsection*{Random null and identical systems models}
It is useful to compare the value of the Joint Recurrence Coupling Indicator (JRCI) to known analytical models that have a straightforward interpretation. Here, we present two such models: the random null model and the identical systems model.

\begin{figure}
    \centering
    \includegraphics[width=0.6\linewidth]{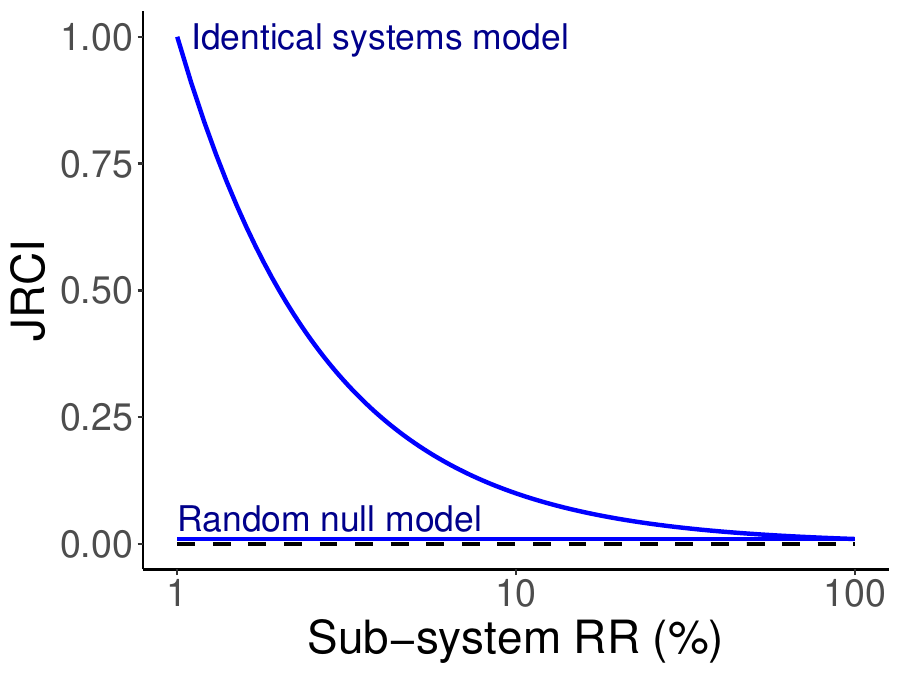}
    \caption{\textbf{JRCI for identical systems model and random null model.} Also shown is the theoretical minimum value of zero (dashed line).}
    \label{fig:theoretical_jrci}
\end{figure}

\paragraph{The identical systems model.} This model is composed of two identical systems, i.e., one system's (multivariate) time series is an exact copy of the other system's time series. In this trivial case the two systems ($x$ and $y$) will have identical recurrence plots and therefore the joint recurrence plot will be identical to each of the recurrence plots, since every recurrence is shared by both systems and is therefore also a joint recurrence (cf.\ Equation~\ref{eq:J}). In other words:
\begin{equation}
    \mathbf{J}_\text{identical} = \mathbf{R}_x \circ \mathbf{R}_y
        = \mathbf{R}_x \circ \mathbf{R}_x
        = \mathbf{R}_y \circ \mathbf{R}_y
        = \mathbf{R}_x = \mathbf{R}_y.
\end{equation}
It follows that $JRR = RR$, where $RR$ is the recurrence rate of both systems. This gives the joint recurrence coupling indicator:
\begin{equation}
    \text{JRCI}_\text{identical} =  \frac{\text{JRR}}{RR^2} = \frac{1}{RR}.
\end{equation}
This expression, shown as the upper curve in Figure~\ref{fig:theoretical_jrci}, diverges as $RR \to 0$ and when the recurrence rate is expressed in percent it has the value $0.01$ when $RR = 100\%$.

\paragraph{The random null model.} This model consists of two systems whose recurrences are distributed so that the recurrence plot of each of the systems is distributed according to i.i.d. Bernoulli random variables\footnote{Or, equivalently, a binomial distribution with a single trial.} on a 2D grid. Each element of the recurrence matrix has a probability $p$ of the value 1 (a recurrence) and a probability $1-p$ of having the value 0 (not a recurrence). In our case we are interested in two systems with the same recurrence rate $RR$, so we set $p = RR$. Since the joint recurrence plot is the element-wise product, each element of the joint recurrence matrix will have a probability $p^2 = RR^2$ of being a recurrence and a probability $1 - p^2$ of non-recurrence, so the expectation value for JRR is $RR^2$.This gives
\begin{equation}
    \text{JRCI}_\text{random} = \frac{\text{JRR}}{RR^2} = 1,
\end{equation}
where we have implicitly taken the recurrence rate to be a fraction, since we set $p = RR$. If we express the recurrence rates in percent, we get:
\begin{equation}
    \text{JRCI}_\text{\%,random} =  \frac{\text{JRR}}{RR^2} 
    =  \frac{100\% \cdot RR^2}{(100\% \cdot RR)^2} = 0.01.
\end{equation}
This is shown as the solid horizontal line in Figure~\ref{fig:theoretical_jrci}.

\subsubsection*{The effect of noise}
All measurements of real physical systems have some level of measurement uncertainty, which we will refer to as measurement noise and in most cases a system is also perturbed by its environment, which we refer to as dynamical noise \cite{szpiro_measuring_1993, monster_causal_2017}.  The main difference is that the system itself is unaffected by measurement noise, whereas dynamical noise affects one or more of the variables in the system and therefore also the system's dynamics.

To examine the effect of measurement noise we take a time series from the Lorenz attractor and add measurement noise at varying noise levels, modeled as Gaussian noise with zero mean and a standard deviation $\sigma_\text{noise} = \xi \sigma_\text{signal}$, and refer to $\xi$ as the noise level. We use the Lorenz system and take one realized time series and add measurement noise to each of the three variables with the same noise level, $\xi$, so that the noise level is the same in units of the standard deviation of the signal.

\begin{figure}
    \centering
    \includegraphics[width=0.6\linewidth]{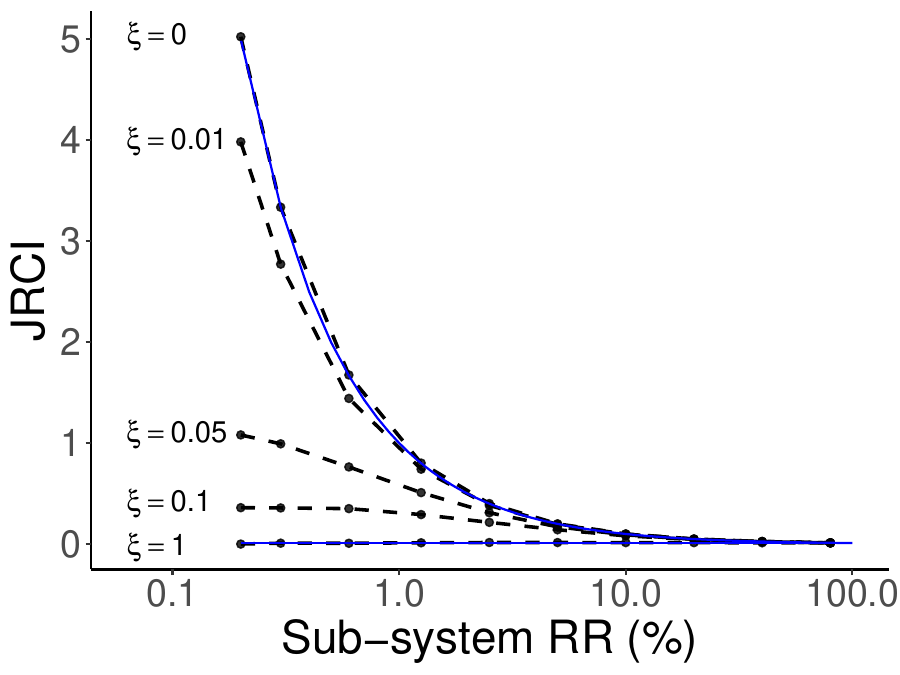}
    \caption{\textbf{The effect of measurement noise.} With no noise $(\xi = 0)$ the results are consistent with the identical systems model, as expected. Increasing values of noise produce larger deviations from the identical systems model, and for $\xi = 1$ the results are consistent with the random null model.}
    \label{fig:lorenz-measurement-noise}
\end{figure}

The results are shown in Figure~\ref{fig:lorenz-measurement-noise}, where two identical Lorenz systems without any measurement noise $(\xi = 0)$ coincide with the identical systems model, as expected. Increasing levels of noise result in lower values of JRCI  and when the variance of the noise equals that of the signal $(\xi = 1)$ the resulting values of JRCI are consistent with the random null model. Note, that there is only a single realization of the model system for each value of $\xi$, so there are no error bars in the plot.

\subsubsection*{High-dimensional Lorenz 96 system and oscillator}
\begin{figure}
    \centering
    \includegraphics[width=\linewidth]{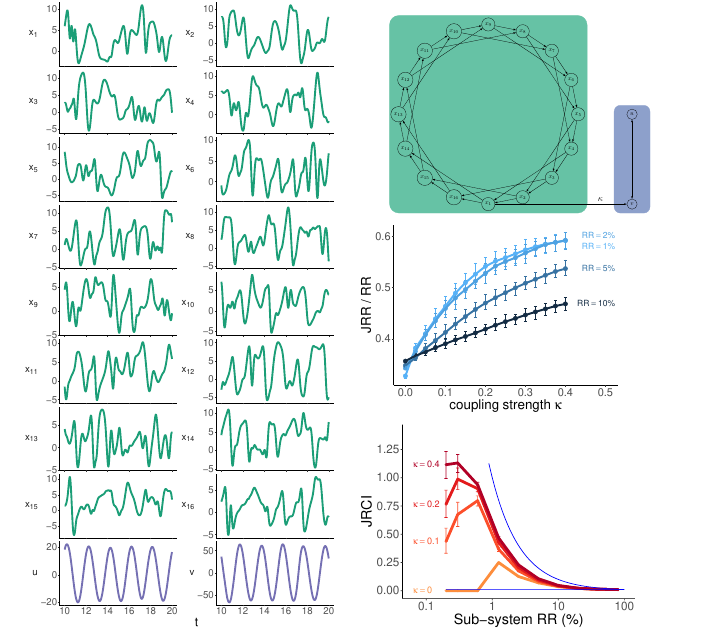}
    \caption{\textbf{Lorenz 96 system coupled to harmonic oscillator.} One component ($x_1$) of a 16-dimensional Lorenz 96 system is coupled uni-directionally to the velocity ($v$) of a harmonic oscillator. Conceptual model (top-right); time series (left); $JRR/RR^2$ as a function of coupling, $\kappa$, for different values of subsystem $RR$ (center-right); JRCI plot for different values of the coupling, $\kappa$ (bottom-right). } 
    \label{fig:lor96hidim}
\end{figure}

In order to test MvJRQA on a synthetic system with higher dimensionality than the relatively low-dimensional systems included, we present results for a Lorenz 96 system with $K=16$ coupled to a harmonic oscillator. This model has a larger difference in dimensionality between the two coupled systems and is closer in this regard to the empirical dataset. The equations for the Lorenz 96 system are given in Equation~\ref{eq:lorenz96} and the harmonic oscillator driven by the $x_1$-variable of the Lorenz~96 system is described in Equation~\ref{eq:osc_l96}. The model is shown, conceptually, in Figure~\ref{fig:lor96hidim}, along with example time series. As for the other model systems, we fixed the coupling strength, $\kappa$, at a series of values and for each value we performed 100 simulations with random initial conditions, skipped the first 100 points, and saved the following 500 points as a time series. We then used the variables $x_1, x_2, \ldots, x_{16}$ from the Lorenz 96 system as coordinates in a 16-dimensional phase space and used the variables $u, v$ from
the harmonic oscillator as coordinates in a 2-dimensional phase space. For each of these two trajectories, we applied MdRQA to obtain the recurrence plots of the two systems at different, fixed, values of the recurrence rate and used these to construct the JRP, i.e. we performed MvJRQA with fixed $RR$. This enabled us to plot §JRR/RR§ and JRCI as shown in Figure~\ref{fig:lor96hidim}. The results are seen to be very similar to those with $K = 5$, shown in panel D of Figure~\ref{fig:results-overview}. In other words, in this particular example, increasing the number of variables in the Lorenz 96 system by more than a factor of 3 did not qualitatively change the results from applying MvJRQA across a wide range of coupling strengths and fixed subsystem recurrence rates.

\subsubsection*{Sensitivity of empirical data analysis to subsystem recurrence rate}
\begin{figure}
    \centering
    \includegraphics[width=0.95\linewidth]{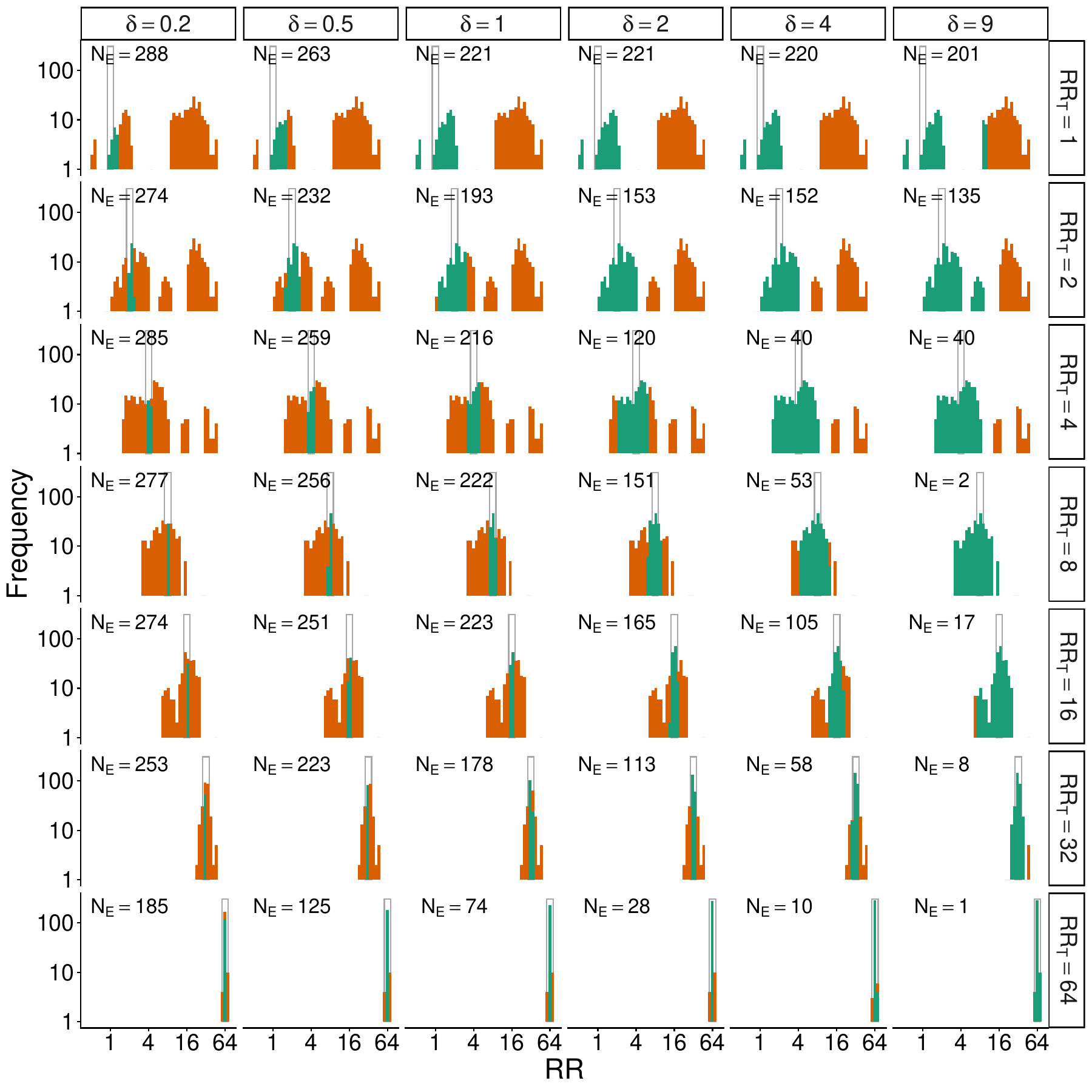}
    \caption{\textbf{Inability to fix $\mathbf{RR}$ at target value for 2D eye tracking.} Each panel shows the histogram of obtained recurrence rates, $RR$ computed over all the 2D eye tracking time series in the empirical dataset. The panels are organized by target recurrence rate, $RR_T$ (rows) and by the tolerance parameter $\delta$ (columns). The ideal histogram, shown in white with gray border, has a single mode at $RR = RR_T$, and the observed deviation from this is due to the inability to find a radius that gives the target recurrence rate, $RR_T$. Observations with $|RR - RR_T| > \delta$, to be excluded,l are shown in red and the number, $N_E$, of these is given for each panel. Note that both axes are logarithmic.}
    \label{fig:delta_hist}
\end{figure}

One crucial aspect of MvJRQA is setting the radius parameter, $\varepsilon$ (in Equation~\ref{eq:RR_def}) so that the recurrence rate of each system or subsystem has a fixed target value: $RR_T$. This can sometimes be impossible to achieve exactly or even to within a desired approximate value. For the empirical example data this is the case for the 2D eye tracking data, where $RR$ is not a continuous function of the radius parameter, $\varepsilon$, and $RR$ increases in discontinuous steps with increasing $\varepsilon$. This makes it impossible to find a value of $\varepsilon$ that fixes the recurrence rate at low values of $RR_T$. Hence, there is a need to remove observations where the obtained value of $RR$ is too different from the desired value, $RR_T$. We operationalize this by introducing a tolerance parameter, $\delta$, such that observations where $RR$ deviates from $RR_T$ by more than $\delta$ are excluded from the analysis. Thus, our exclusion criterion is:
\begin{equation}\label{eq:exclusion}
    |RR - RR_T| > \delta.
\end{equation}
We can then investigate how the choice of $\delta$ influences the results and, as a first step, how many observations are excluded for particular values of $RR_T$ and $\delta$. A visual overview of the distribution of $RR$ and the number of excluded observations for different values of $RR_T$ and $\delta$ is shown in Figure~\ref{fig:delta_hist}. For low target recurrence, $RR_T$, and low values of the tolerance parameter, $\delta$ more observations are excluded; whereas fewer observations are excluded for higher values of $RR_T$ and $\delta$. We therefore face a pragmatic need to choose a value of $\delta$ that is not so low that most observations are excluded and not so high that we contaminate the data by including observations with values of $RR$ that are too far from $RR_T$. A value in the range $1 \leq \delta \leq 2$ is seen to strike a balance between these two opposing needs.

\begin{figure}
    \centering
    \includegraphics[width=0.9\linewidth]{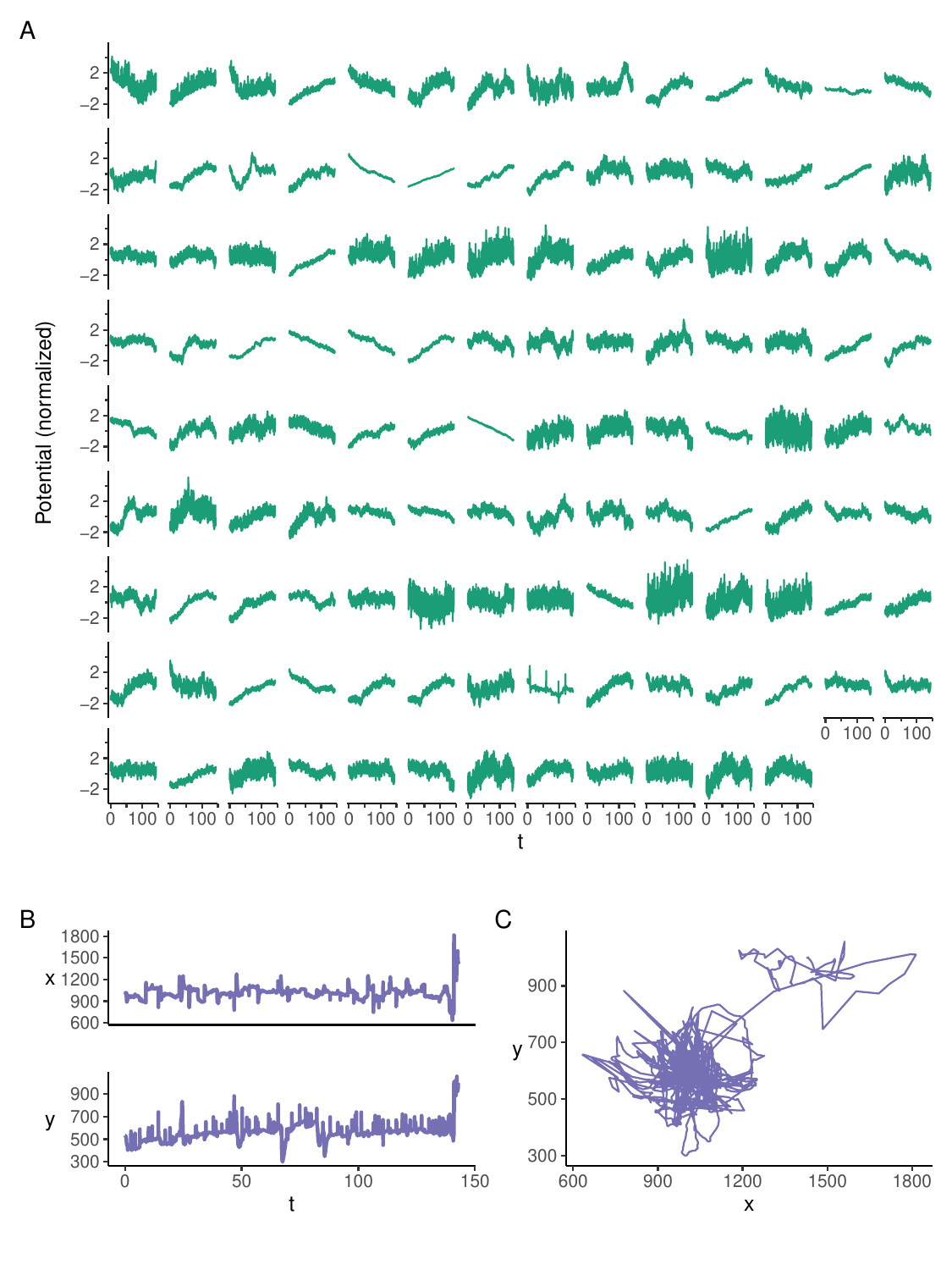}
    \caption{\textbf{Example empirical time series data.} Normalized (z-scored) time series for 124 channels of EEG data (\textbf{A}) and $x$ and $y$ components of eye tracking (\textbf{B}) for a single experimental trial.  Panel \textbf{C} shows the eye tracking trajectory.}
    \label{fig:empirical-timeseries}
\end{figure}
 
\begin{figure}
    \centering
    \includegraphics[width=0.95\linewidth]{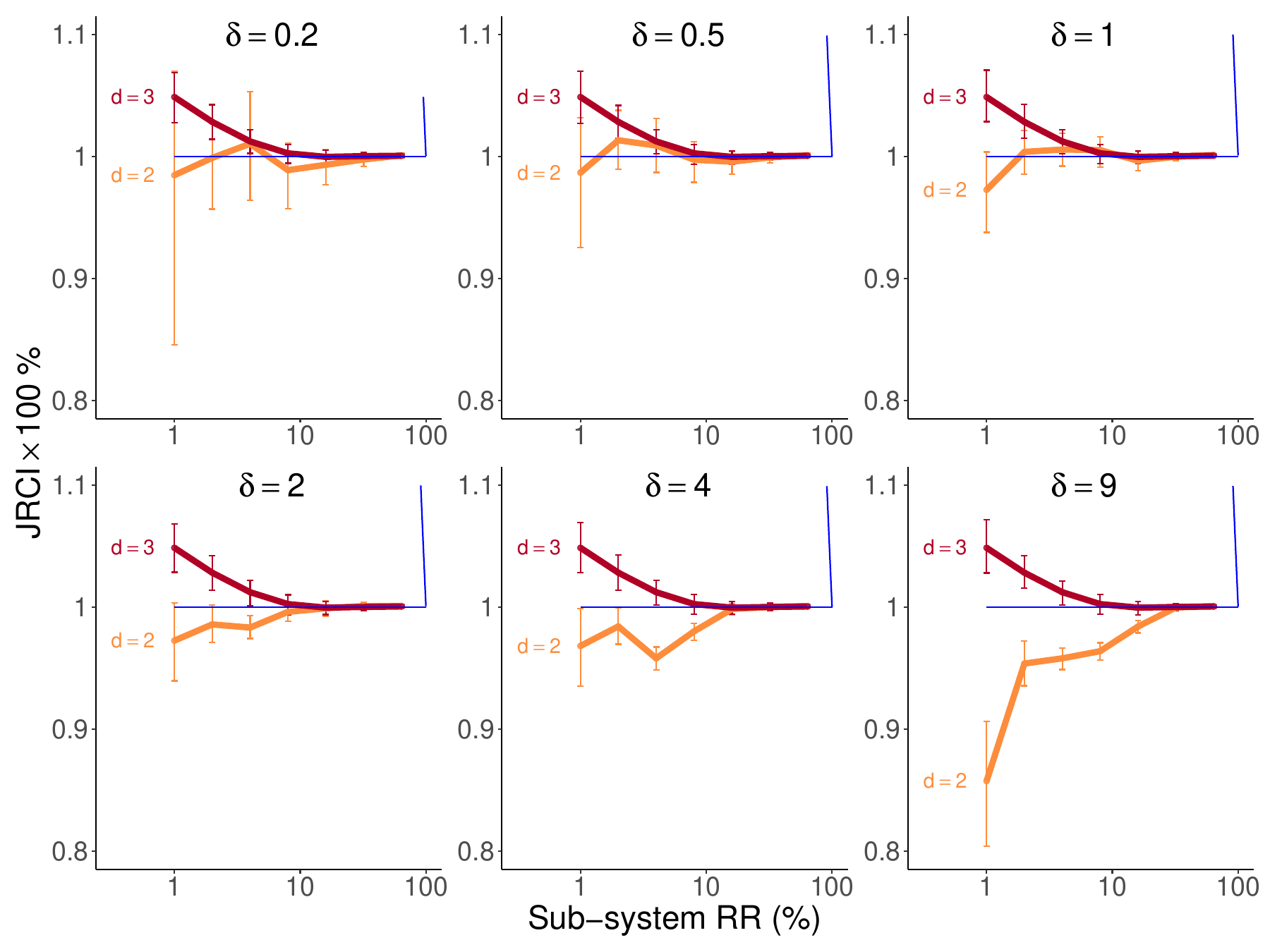}
    \caption{\textbf{Effect of tolerance parameter on joint recurrence coupling indicator.} Each panel shows a plot of the joint recurrence coupling indicator, JRCI ($JRR/RR^2$) as a function of fixed subsystem recurrence rate, $RR$, for both 2D ($d = 2$) and 3D ($d = 3$) eye tracking as one of the subsystems and EEG as the other subsystem. The panels differ in the value of the tolerance parameter, $\delta$, which is used to exclude observations according to Equation~\ref{eq:exclusion} as shown in Figure~\ref{fig:delta_hist}. Error bars show bootstrapped 95\% CI. The blue horizontal line is the random null model.} 
    \label{fig:jrci_delta}
\end{figure}

For each of the values of the tolerance parameter, $\delta$, the observations remaining after applying the exclusion criterion in Equation~\ref{eq:exclusion} can be used to construct a plot of the joint recurrence coupling indicator, JRCI, as shown in Figure~\ref{fig:jrci_delta}. Except for the lowest value of $\delta$, no observations are excluded from the EEG and the 3D eye tracking data, so the main effect is on the 2D eye tracking where a large fraction of the observations are excluded for low to moderate values of $\delta$ and at low target recurrence rate, $RR_T$, even at the highest value of $\delta = 10$.

For $\delta = 0.1$ most observations for $d = 2$ are excluded, resulting in very large 95\% confidence intervals. At moderate values, $0.5 \leq \delta \leq 4$, more observations are included, resulting in smaller error bars and the results for EEG and 2D eye tracking ($d = 2$ in Figure~\ref{fig:jrci_delta}) are consistent with the random null model, indicating no coupling. For the largest value of the tolerance parameter, $\delta = 10$, some observations with $RR$ very far from $RR_T$ are included which has the effect of spuriously suppressing JRCI below the random null model. This is already evident at $\delta = 4$, indicating that a lower value is more appropriate. We note that a plot of JRCI that is consistently below the random null model is a sign that the algorithm has not been able to fix subsystem recurrence at values close to $RR_T$ and for this reason the supplied implementation includes the parameter \texttt{tolerance} in the function \texttt{find\_threshold()} with a default value of 0.5. It is important to note that exceeding the chosen level of tolerance will only result in a warning message, and it is up to the user to further investigate the results and determine if observations should be excluded, e.g., in a similar way to how we have described it here. 

Based on the data presented in Figure~\ref{fig:delta_hist} and \ref{fig:jrci_delta} we conclude that for this particular dataset---and specifically the 2D eye tracking data---the most appropriate value of the tolerance parameter is in the range $1 \leq \delta \leq 2$ and hence, we use $\delta = 1.5$ in the regression analysis (Table~\ref{tab:regression}) and the JRCI plot (Figure~\ref{fig:empirical-jrci}).
 
\begin{figure}
    \centering
    \includegraphics[width=0.6\linewidth]{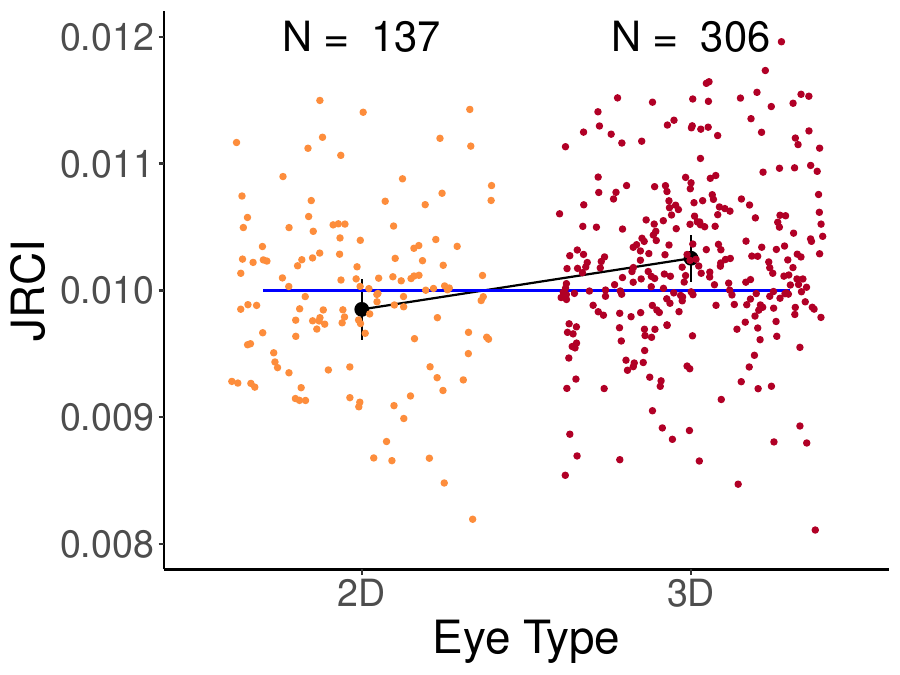}
    \caption{\textbf{Regression model and data.} Individual points show the JRCI-values computed with MvJRQA from a single experimental trial with target recurrence rate $RR_T = 2\%$. For 3D eye tracking data all 306 trials were included, but for 2D eye tracking 169 observations were removed, since it was not possible to fix $RR$ in the interval $[0.5\%, 3.5\%]$, i.e. $\delta = 1.5\%$ and $RR_T = 2\%$, resulting in 137 observations in the interval $RR_T \pm \delta$. The black points and error bars are marginal means with 95\% CI for the regression model in Equation~\ref{eq:mlm} whose results are shown in Table~\ref{tab:regression}.}
    \label{fig:regression}
\end{figure}
 
As a robustness check, to ensure that our results are not an artefact of our choice of $\delta$, we perform two extra regression analyses, also using Equation~\ref{eq:mlm} as for the results in Table~\ref{tab:regression}. In model 1 (see Table~\ref{tab:robustness}), we have included all observations, irrespective of the distance between $RR$ and $RR_T$, i.e. corresponding to $\delta = 100$. In model 2, we kept $\delta = 1.5$, as in the main analysis, and removed not only the observations for EEG and 2D eye tracking using the exclusion criterion in Equation~\ref{eq:exclusion} but also the observations for EEG and 3D eye tracking. These two models, identical in structure to the model presented in Table~\ref{tab:regression}, represent two opposite extremes: removing no observations at all (model 1) and removing observations for both 2D and 3D eye tracking (model 2). Thus, these two models, together, are a robustness check on the exclusion criterion applied in the main analysis ($\delta = 1.5$). Comparing the results in Table~\ref{tab:regression} with those in Table~\ref{tab:robustness}, we see that the overall conclusion is the same based on all three models, but with bigger effect sizes in the robustness checks in the models in Table~\ref{tab:robustness}.

\begin{table}
\caption{\textbf{Robustness check on regression results for MvJRQA.} Model 1 is the same model shown in Table~\ref{tab:regression}, but including all 612 observations, i.e., no, observations were removed, even if $RR - RR_T > \delta$. Model 2 represents the other extreme, viz.\ removing the MvJRQA results for EEG combined with both the 2D and the 3D eye tracking results in all the cases where $RR - RR_T > \delta$ for 2D eye tracking.}
\begin{center}
\resizebox{\textwidth}{!}{%
\begin{tabular}{l c c c c c c c c}
\hline
 & \multicolumn{4}{c}{Model 1} & \multicolumn{4}{c}{Model 2} \\
\cline{2-5} \cline{6-9}
 & Estimate & SE & $t$ & $p$ & Estimate & SE & $t$ & $p$ \\
\hline
(Intercept)    & $-0.59$           & $0.09$ & $-6.25$ & $<0.001$ & $-0.24$           & $0.11$ & $-2.12$ & $0.035$ \\
               & $ [-0.77; -0.40]$ & $$     & $$      & $$     & $ [-0.45; -0.02]$ & $$     & $$      & $$     \\
EyeType3D      & $1.15$            & $0.06$ & $20.34$ & $<0.001$ & $0.38$            & $0.11$ & $3.32$  & $0.001$ \\
               & $ [ 1.04;  1.26]$ & $$     & $$      & $$     & $ [ 0.15;  0.60]$ & $$     & $$      & $$     \\
\hline
Num. obs.      & $612$             & $$     & $$      & $$     & $274$             & $$     & $$      & $$     \\
Conditional $R^2$ & $0.51$            & $$     & $$      & $$     & $0.13$            & $$     & $$      & $$     \\
\hline
\end{tabular}
}
\label{tab:robustness}
\end{center}
\end{table}

\subsubsection*{Replication script and software}
A collection of scripts in \textsc{R} \cite{r_core_team_r_nodate} using renv \cite{ushey_renv_2025} to reproduce all the results in the article is available from Zenodo \cite{monster_mvjrqa_2025} and GitHub: \url{https://github.com/danm0nster/mvjrqa-replication}.

The scripts make use of the \textsc{R} packages:
\textsc{animation} \cite{xie_animation_2013},
\textsc{cowplot} \cite{wilke_cowplot_2025},
\textsc{crqa} \cite{coco_r_2021},
\textsc{data.table} \cite{barrett_datatable_2025},
\textsc{deSolve} \cite{soetaert_solving_2010},
\textsc{dplyr} \cite{wickham_dplyr_2023},
\textsc{edf} \cite{henelius_edf_2016},
\textsc{effectsize} \cite{ben-shachar_effectsize_2020},
\textsc{gridExtra} \cite{auguie_gridextra_2017},
\textsc{ggplot2} \cite{wickham_ggplot2_2016},
\textsc{ggrepel} \cite{slowikowski_ggrepel_2024},
\textsc{latex2exp} \cite{meschiari_latex2exp_2022},
\textsc{lme4} \cite{bates_fitting_2015},
\textsc{lmerTest} \cite{kuznetsova_lmertest_2017},
\textsc{Matrix} \cite{bates_matrix_2025},
\textsc{modelbased} \cite{makowski_modelbased_2025},
\textsc{parallel} \cite{r_core_team_r_nodate},
\textsc{patchwork} \cite{pedersen_patchwork_2025},
\textsc{performance} \cite{ludecke_performance_2021},
\textsc{purrr} \cite{wickham_purrr_2025},
\textsc{readr} \cite{wickham_readr_2024},
\textsc{rgl} \cite{murdoch_rgl_2025},
\textsc{see} \cite{ludecke_see_2021},
\textsc{texreg} \cite{leifeld_texreg_2013}, and
\textsc{tidyr} \cite{wickham_tidyr_2024}.

\begin{figure}
    \centering
    \includegraphics[width=0.8\linewidth]{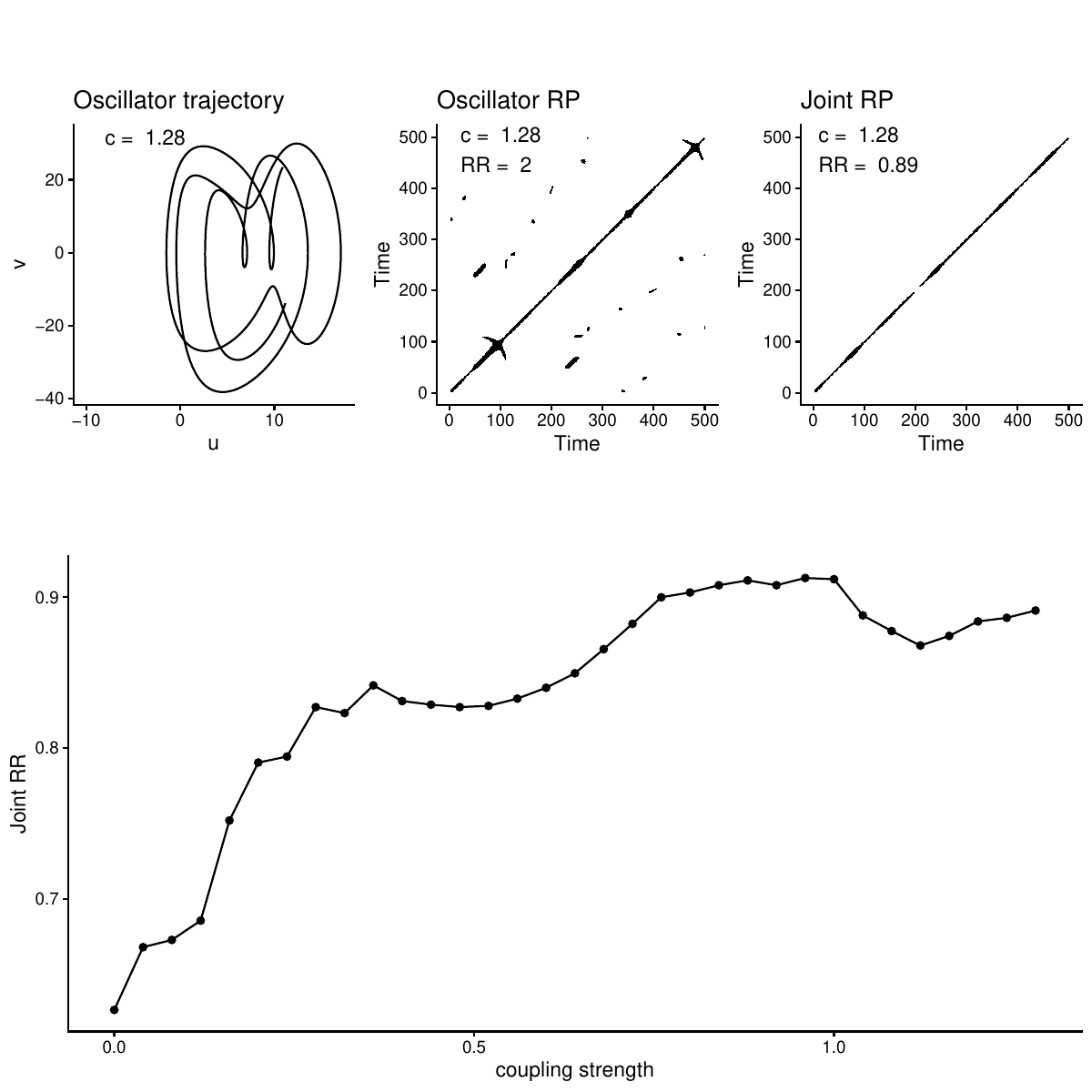}
    \caption{\textbf{Final frame of Movie S1.} Each frame of the movie shows the effect of coupling for a particular value of the coupling, starting at $c = 0$. The trajectory of the driven harmonic oscillator system is shown in the upper left panel.  The recurrence plot of the harmonic oscillator at fixed recurrence rate $(RR \approx 2\%)$ is shown in the upper middle panel. The joint recurrence plot is shown in the upper right panel. The bottom panel is a plot that shows the joint recurrence rate, $JRR$, for each value of the coupling, and is built up frame-by-frame as the movie plays. The upper left panel in the movie is an animated version of Figure~\ref{fig:harmonic_ps}, extended to extreme values of the coupling, $c$.}
    \label{fig:movie_snapshot}
\end{figure}

\clearpage %

\paragraph{Caption for Movie S1.}
\textbf{Effect of unidirectional coupling on driven system's recurrence plot and joint recurrence plot.} The movie extends Figure~\ref{fig:harmonic_ps} to show how increasing the coupling, $c$, affects the harmonic oscillator's trajectory in phase space, with increasing deviation from the unperturbed trajectory. It also shows how this changes the recurrence plot of the perturbed harmonic oscillator at fixed subsystem recurrence rate and, in turn, how this changes the joint recurrence plot and one of its main characteristics, the joint recurrence rate, $JRR$. A version of this plot based on a larger sample of random initial conditions is shown in Figure~\ref{fig:extreme-coupling}C. Not shown in the movie are the trajectory of the Lorenz system and its recurrence plot, but these are unchanged throughout, since the coupling is unidirectional from the Lorenz system to the harmonic oscillator, cf. Equation~\ref{eq:lh_model}. Only a single realization of the system for each value of $c$ is shown. The layout of the panels in the movie is explained in figure~\ref{fig:movie_snapshot} which is a snapshot of the final frame of the movie. The initial state is $x = 10$, $y = 10$, $z = 10$,
$u = 10$, $v = 0$, and the first 500 points are discarded as transient, and the following 500 points are used in the time series.

\end{document}